\pdfoutput=1
\documentclass[11pt,twoside,a4paper,cmspaper,final,collab]{cms-tdr}
\def\svnVersion{6f4f302-D}\def\svnDate{2020/09/29}\def\cmsCernNoTag{CERN-EP-2020-131}\def\cmsCernDate{\today}\def\cmsMessage{"Published in the European Physical Journal C as \href{http://dx.doi.org/10.1140/epjc/s10052-020-08701-5}{\doi{10.1140/epjc/s10052-020-08701-5}.}"}
\begin{document}\cmsNoteHeader{SUS-19-011}

\ifthenelse{\boolean{cms@external}}{\providecommand{\cmsLeft}{upper\xspace}}{\providecommand{\cmsLeft}{left\xspace}}
\ifthenelse{\boolean{cms@external}}{\providecommand{\cmsRight}{lower\xspace}}{\providecommand{\cmsRight}{right\xspace}}

\newlength\cmsTabSkip\setlength{\cmsTabSkip}{1ex}

\newcommand{\Njets}{\ensuremath{N_\text{jets}}\xspace}
\newcommand{\Nbjets}{\ensuremath{N_{\PQb}}\xspace}

\newcommand{\Ttt}{\ensuremath{\text{T2}\PQt\PQt}\xspace}
\newcommand{\TbW}{\ensuremath{\text{T2}{\PQb\PW}}\xspace}
\newcommand{\Tbbllnunu}{\ensuremath{\text{T8}\PQb\PQb\ell\ell\PGn\PGn}\xspace}

\newcommand{\ttZ}{\ensuremath{\ttbar\cPZ}\xspace}
\newcommand{\tqZ}{\ensuremath{\PQt\PQq\cPZ}\xspace}
\newcommand{\ttW}{\ensuremath{\ttbar\PW}\xspace}
\newcommand{\ttH}{\ensuremath{\ttbar\PH}\xspace}
\newcommand{\VVV}{\ensuremath{\PV\PV\PV}\xspace}
\newcommand{\VV}{\ensuremath{\PV\PV}\xspace}
\newcommand{\tZq}{\ensuremath{\cPqt\PZ\Pq}\xspace}
\newcommand{\ttgamma}{\ensuremath{\ttbar\Pgg^{(*)}}\xspace}
\newcommand{\tWZ}{\ensuremath{\cPqt\PW\PZ}\xspace}
\newcommand{\tHq}{\ensuremath{\cPqt\PH\Pq}\xspace}
\newcommand{\tHW}{\ensuremath{\cPqt\PH\PW}\xspace}

\newcommand{\pp}{\ensuremath{\Pp\Pp}\xspace}
\newcommand{\ptmissSig}{\ensuremath{\mathcal{S}}\xspace}
\newcommand{\metSig}{\ptmissSig}
\newcommand{\mt}{\ensuremath{M_{\text{T}}}\xspace}
\newcommand{\mtll}{\ensuremath{M_{\text{T2}}(\ell \ell)}\xspace}
\newcommand{\mtlblb}{\ensuremath{M_{\text{T2}}(\PQb\ell \PQb\ell )}\xspace}
\newcommand{\lumiGolden}{137\fbinv}
\providecommand{\CL}{CL\xspace}
\providecommand{\cmsTable}[1]{\resizebox{\textwidth}{!}{#1}}

\cmsNoteHeader{SUS-19-011}
\title{Search for top squark pair production using dilepton final states in \texorpdfstring{\pp}{pp} collision data collected at \texorpdfstring{$\sqrt{s}=13\TeV$}{sqrt(s) = 13 TeV}}

\date{\today}

\abstract{
A search is presented for supersymmetric partners of the top quark (top squarks) in final states with two oppositely charged leptons (electrons or muons), jets identified as originating from \PQb quarks, and missing transverse momentum.
The search uses data from proton-proton collisions at $\sqrt{s}=13\TeV$ collected with the CMS detector, corresponding to an integrated luminosity of \lumiGolden.
Hypothetical signal events are efficiently separated from the dominant top quark pair production background with requirements on the significance of the missing transverse momentum and on transverse mass variables.
No significant deviation is observed from the expected background.
Exclusion limits are set in the context of simplified supersymmetric models with pair-produced lightest top squarks.
For top squarks decaying exclusively to a top quark and a lightest neutralino, lower limits are placed at $95\%$ confidence level on the masses of the top squark and the neutralino up to 925 and 450\GeV, respectively.
If the decay proceeds via an intermediate chargino, the corresponding lower limits on the mass of the lightest top squark are set up to 850\GeV for neutralino masses below 420\GeV.
For top squarks undergoing a cascade decay through charginos and sleptons, the mass limits reach up to 1.4\TeV and 900\GeV respectively for the top squark and the lightest neutralino.
}

\hypersetup{%
pdfauthor={CMS Collaboration},%
pdftitle={Search for top squark pair production in the dilepton final state using pp collision data collected at sqrt(s) = 13 TeV with the CMS experiment},%
pdfsubject={CMS},%
pdfkeywords={CMS, supersymmetry, top squark}}

\maketitle

\section{Introduction}
\label{sec:Introduction}
The standard model (SM) of particle physics accurately describes the overwhelming majority of observed particle physics phenomena.
Nevertheless, several open questions are not addressed by the SM, such as the hierarchy problem, the need for fine tuning to reconcile the
large difference between the electroweak and the Planck scales in the presence of a fundamental scalar~\cite{Witten:1981nf,Barbieri:1987fn,Aad:2012tfa,Chatrchyan:2012ufa}.
Moreover, there is a lack of an SM candidate particle that could constitute the dark matter in cosmological and astrophysical observations~\cite{Bertone:2004pz,Feng:2010gw}.
Supersymmetry (SUSY)~\cite{Ramond:1971gb,Golfand:1971iw,Neveu:1971rx,Volkov:1972jx,Wess:1973kz,Wess:1974tw,Fayet:1974pd,Nilles:1983ge} is a well-motivated extension of the SM that provides a solution to both of these problems, through the introduction
of a symmetry between bosons and fermions.
In SUSY models, large quantum loop corrections to the mass of the Higgs boson (\PH), mainly arising from the top quarks, are mostly
canceled by those arising from their SUSY partners, the top squarks, if the masses of the SM particles and their SUSY partners are close in value.
Similar cancellations occur for other particles, resulting in a natural solution to the
hierarchy problem~\cite{Barbieri:1987fn,Dimopoulos:1981zb,Kaul:1981hi}. Furthermore, SUSY introduces a new quantum number, R~parity~\cite{Farrar:1978xj}, that
distinguishes between SUSY and SM particles. If R~parity is conserved, top squarks are
produced in pairs and the lightest SUSY particle~(LSP) is stable. If neutral, the LSP provides a good
candidate for the dark matter.
The lighter top squark mass eigenstate \PSQtDo is the lightest squark in many SUSY models and may be within the energy reach of the CERN LHC if SUSY provides a natural solution to the hierarchy problem~\cite{Papucci:2011wy}.
This strongly motivates searches for top squark production.

In this paper, we present a search for top squark pair production in data from proton-proton~(\pp) collisions collected at a center-of-mass energy of 13\TeV, corresponding to an integrated luminosity of \lumiGolden, with the CMS detector at the LHC from 2016 to 2018.
The search is performed in final states with two leptons (electrons or muons), hadronic jets identified as originating from \PQb quarks, and significant missing transverse momentum (\ptmiss).
The large background from the SM top quark-antiquark pair production (\ttbar) is reduced by several orders of magnitude through the use of specially designed transverse-mass variables~\cite{Smith:1983aa,Lester:1999tx}.
Simulations of residual SM backgrounds in the search regions are validated in control regions orthogonal to the signal regions, using observed data.

\begin{figure*}[htb]
\centering
\includegraphics[width=0.32 \textwidth]{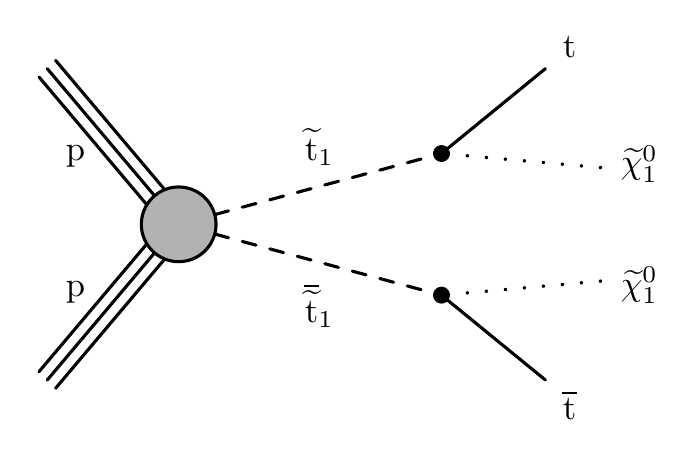}
\includegraphics[width=0.32 \textwidth]{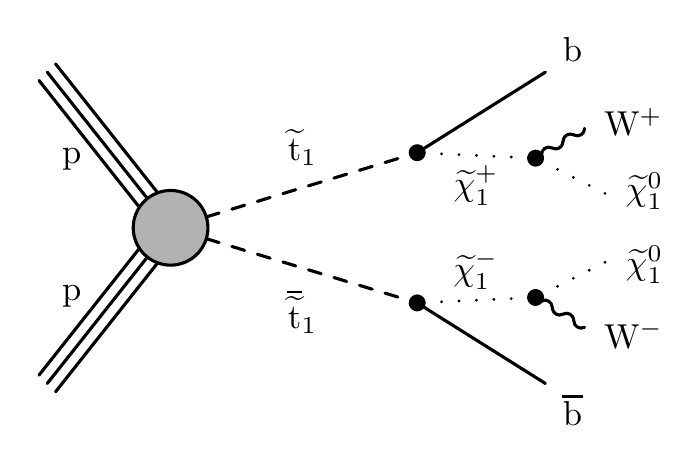}
\includegraphics[width=0.32 \textwidth]{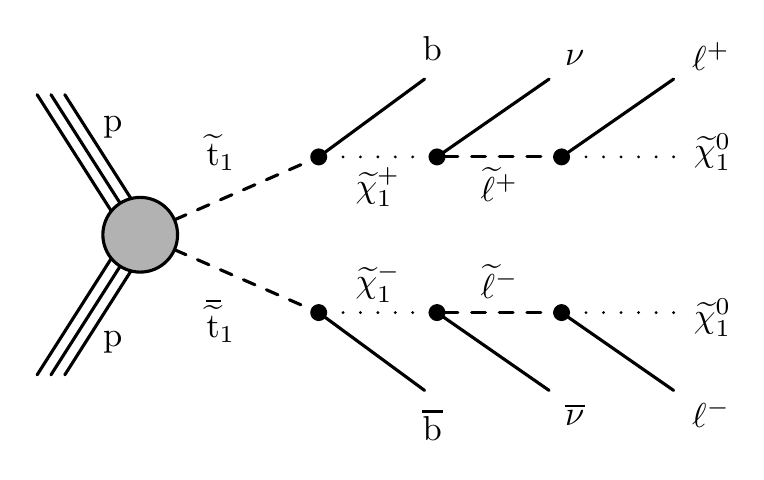}
\caption{
Diagrams for simplified SUSY models with strong production of top squark pairs $\PSQtDo\PASQt_{1}$.
In the \Ttt model~(left), the top squark decays to a top quark and a \PSGczDo.
In the \TbW model~(center), the top squark decays into a bottom quark and an intermediate \PSGcpmDo that further decays into a \PW boson and a \PSGczDo.
The decay of the intermediate \PSGcpmDo, which yields a \PGn, plus a \PSGczDo and a $\ell^\pm$ from the decay of an intermediate slepton $\widetilde{\ell}^{\pm}$, is described by the \Tbbllnunu model~(right).
}
\label{fig:fey}
\end{figure*}

Simplified models~\cite{Alwall:2008ag,Alwall:2008va,Alves:2011wf} of strong top squark pair production and different top squark decay modes are considered.
Following the naming convention in Ref.~\cite{Chatrchyan:2013sza}, top squark decays to top quarks and neutralinos~(\PSGczDo, identified as LSPs) are described by the \Ttt model (Fig.~\ref{fig:fey}, left).
In the \TbW model (Fig.~\ref{fig:fey}, center), both top squarks decay via an intermediate chargino (\PSGcpmDo) into a bottom quark, a \PW~boson, and an LSP.
In both models, the undetected LSPs and the neutrinos from leptonic \PW decays account for significant \ptmiss, and the leptons provide a final state with low SM backgrounds.
In the \Tbbllnunu model (Fig.~\ref{fig:fey}, right), both top squarks decay via an intermediate chargino to a bottom quark, a slepton, and a neutrino.
The branching fraction of the chargino to sleptons is assumed to be identical for the three slepton flavors.
The subsequent decay of the sleptons to neutralinos and leptons leads to a final state with the same particle content as in the \Ttt model, albeit without the suppression of the dilepton final state from the leptonic \PW boson branching fraction.

Searches for top squark production have been performed by the ATLAS ~\cite{Aad:2015pfx,Aad:2014kra,Aad:2014qaa,Aaboud:2016lwz,Aaboud:2017nfd,Aaboud:2017ayj,Aaboud:2017aeu,Aad:2020sgw} and CMS~\cite{Chatrchyan:2013xna,Khachatryan:2016pup,Sirunyan:2016jpr,Sirunyan:2017xse,Sirunyan:2017wif,Sirunyan:2017leh,Sirunyan:2019zyu,Sirunyan:2019glc} Collaborations using 8 and 13\TeV \pp collision data.
These searches disfavor top squark masses below about 1.1--1.3\TeV in a wide variety of production and decay scenarios.
Here we present a search for top squark pair production in dilepton final states.
With respect to a previous search in this final state~\cite{Sirunyan:2017leh}, improved methods to suppress and estimate backgrounds from SM processes and a factor of about four larger data set increase the expected sensitivity by about 125\GeV in the \PSQtDo mass.
This search complements recent searches for top squark production in other final states~\cite{Sirunyan:2019zyu,Sirunyan:2019glc}, in particular in scenarios with a compressed mass spectrum or final states with a single lepton.

\section{The CMS detector}
\label{sec:CMS}
The central feature of the CMS apparatus is a superconducting solenoid of 6\unit{m} internal diameter, providing a magnetic field of 3.8\unit{T}.
Within the solenoid volume are a silicon pixel and strip tracker, a lead tungstate crystal electromagnetic calorimeter (ECAL), and a brass and scintillator hadron calorimeter, each composed of a barrel and two endcap sections.
Forward calorimeters extend the pseudorapidity ($\eta$) coverage provided by the barrel and endcap detectors that improve the measurement of the imbalance in transverse momentum.
Muons are detected in gas-ionization chambers embedded in the steel flux-return yoke outside the solenoid.

Events of interest are selected using a two-tier trigger system.
The first level, composed of custom hardware processors, uses information from the calorimeters and muon detectors to
select events in a fixed time interval of less than 4\unit{$\mu$s}. The second level, called the high-level
trigger, further decreases the event rate from around 100\unit{kHz} to less than 1\unit{kHz} before data
storage~\cite{Khachatryan:2016bia}.
A more detailed description of the CMS detector, together with a definition of the coordinate system used and the relevant kinematic variables, can be found in Ref.~\cite{Chatrchyan:2008zzk}.

\section{Event samples}
The search is performed in a data set collected by the CMS experiment during the 2016--2018 LHC running periods.
Events are selected online by different trigger algorithms that require the presence of one or two leptons (electrons or muons).
The majority of events are selected with dilepton triggers.
The thresholds of same-flavor (SF) dilepton triggers are 23\GeV (electron) or 17\GeV (muon) on the transverse momentum (\pt) of the leading lepton, and 12\GeV (electron) or 8\GeV (muon) on the subleading lepton \pt.
Triggers for different-flavor (DF) dileptons have thresholds of 23\GeV on the leading lepton \pt, and 12\GeV (electron) or 8\GeV (muon) on the subleading lepton \pt.
Single lepton triggers with a 24\GeV threshold for muons and with a 27\GeV threshold for electrons (32\GeV for electrons in the years 2017 and 2018) improve the selection efficiency.
The efficiency of this online selection is measured using observed events that are independently selected based on the presence of jets and requirements on the \ptmiss.
Typical efficiencies range from 95 to 99\%, depending on the \pt and $\eta$ of the two leptons and are accounted for by corrections applied to simulated events.
\begin{table*}[htb]
\centering
\topcaption{
Event generator and orders of accuracy for each simulated background process.
}
\label{table:samples}
\begin{tabular}{ccccc}
\multirow{2}{*}{Process}  & Cross section     & \multirow{2}{*}{Event generator} & Perturbative \\
                          & normalization     & & order & \\[.2cm] \hline
\ttbar, single~\PQt & NNLO+NNLL & \POWHEG~v2 & NLO \\
$\PQt\PW$ & NNLO & \POWHEG~v1/v2 & NLO \\
\ttH & NLO+NLL & \POWHEG~v2 & NLO \\
Drell--Yan & NNLO& \MGvATNLO& LO\\ [\cmsTabSkip]
\ttZ, \ttW, \tZq, \ttgamma, & \multirow{2}{*}{NLO} & \multirow{2}{*}{\MGvATNLO} & \multirow{2}{*}{NLO} \\
\VVV, \VV & & & \\ [\cmsTabSkip]
\tHW, \tHq & NLO & \MGvATNLO & LO \\
\tWZ & LO & \MGvATNLO & LO
\end{tabular}
\end{table*}

Simulated samples matching the varying conditions for each data taking period are generated using Monte Carlo (MC) techniques.
The \ttbar production and $t$- and $s$-channel single-top-quark background processes are simulated at next-to-leading order (NLO) with the \POWHEG v2~\cite{Nason:2004rx,Frixione:2007vw,Alioli:2010xd,Campbell:2014kua,Alioli:2009je,Frixione:2007nw,Aliev:2010zk,Kant:2014oha} event generator,
and are normalized to next-to-next-to-leading-order (NNLO) cross sections, including soft-gluon resummation at next-to-next-to-leading-logarithmic (NNLL) accuracy~\cite{Czakon:2011xx}.
Events with single top quarks produced in association with \PW bosons ($\PQt\PW$) are simulated with \POWHEG v1~\cite{Re:2010bp} (2016) or \POWHEG v2~(2017--2018), and are normalized to the NNLO cross section~\cite{Kidonakis:2010ux,Kidonakis:2012rm}.
The \ttH process is generated with \POWHEG v2 at NLO~\cite{Hartanto:2015uka}.
Drell--Yan events are generated with up to four extra partons in the matrix element calculations with \MGvATNLO v2.3.3~(2016) and v2.4.2~(2017--2018)~\cite{Alwall:2014hca} at leading order (LO), and the cross section is computed at NNLO~\cite{Gavin:2010az}.
The \ttZ, \ttW, \tZq, \ttgamma, and triboson~(\VVV) processes are generated with \MGvATNLO at NLO.
The cross section of the \ttZ process is computed at NLO in perturbative quantum chromodynamics (QCD) and electroweak accuracy~\cite{Garzelli:2012bn,Frixione:2015zaa}.
The \ttH process is normalized to a cross section calculated at NLO+NLL accuracy~\cite{deFlorian:2016spz}.
The diboson~(\VV) processes are simulated with up to one extra parton in the matrix element calculations, using \MGvATNLO at NLO.
The $\PQt\PW\PZ$, $\PQt\PH\PQq$, and $\PQt\PH\PW$ processes are generated at LO with \MGvATNLO.
These processes are normalized to the most precise available cross sections, corresponding to NLO accuracy in most cases.
A summary of the event samples is provided in Table~\ref{table:samples}.

The event generators are interfaced with \PYTHIA v8.226 (8.230)~\cite{Sjostrand:2014zea} using the CUETP8M1 (CP5) tune~\cite{Skands:2014pea,Khachatryan:2015pea,Sirunyan:2019dfx} for 2016 (2017, 2018) samples to simulate the fragmentation, parton shower, and hadronization of partons in the initial and final states, along with the underlying event.
The NNPDF parton distribution functions (PDFs) at different perturbative orders in QCD are used in v3.0~\cite{Ball:2014uwa} and v3.1~\cite{Ball:2017nwa} for 2016 and 2017--18 samples, respectively.
Double counting of the partons generated with \MGvATNLO and \PYTHIA is removed using the MLM~\cite{Alwall:2007fs} and the \textsc{FxFx}~\cite{Frederix:2012ps} matching schemes for LO and NLO samples, respectively.
The events are subsequently processed with a \GEANTfour-based simulation model~\cite{Agostinelli:2002hh} of the CMS detector.

The SUSY signal samples are generated with \MGvATNLO at LO precision, with up to two extra partons in the matrix element calculations, interfaced with \PYTHIA v8.226 (8.230) using the CUETP8M1 (CP2) tune for 2016 (2017, 2018).
For the \Ttt and \TbW models, the top squark mass is varied from 200 to 1200\GeV and the mass of the LSP is scanned from 1 to 650\GeV.
The mass of the chargino in the \TbW model is assumed to be equal to the mean of the masses of the top squark and the lightest neutralino.
For the \Tbbllnunu model, the top squark mass is varied from 200 to 1600\GeV and the mass of the LSP is scanned from 1 to 1200\GeV.
Similarly to the \TbW model, the mass of the chargino is assumed to be equal to the mean of the top squark and the LSP masses.
For the slepton mass, three values of $x = 0.95$, 0.50, 0.05 are chosen in $m_{\widetilde{\ell}} = x \, (m_{\PSGcpDo} - m_{\PSGczDo}) + m_{\PSGczDo}$.
The production cross sections of signal samples are normalized to approximate NNLO+NNLL accuracy with all other SUSY particles assumed to be heavy and
decoupled~\cite{Beenakker:1996ch,Kulesza:2008jb,Kulesza:2009kq,Beenakker:2009ha,Beenakker:2011fu,Borschensky:2014cia,Beenakker:2011sf,Beenakker:2013mva,Beenakker:2014sma,Beenakker:2016lwe,Beenakker:1997ut,Beenakker:2010nq,Beenakker:2016gmf}.
The simulation of the detector response is performed using the CMS fast detector simulation~\cite{Abdullin:2011zz,Giammanco:2014bza}.

All simulated samples include the effects of additional \pp collisions in the same or adjacent bunch crossings (pileup), and are reweighted according to the observed distribution of the number of interactions per bunch crossing.
An additional correction is applied to account for a mismatch of the simulated samples and the observed distribution of primary vertices in the 2018 running period.

\section{Object and event selection}
\label{sec:objects}
Event reconstruction uses the CMS particle-flow (PF) algorithm~\cite{Sirunyan:2017ulk}, which provides an exclusive set of electron~\cite{Khachatryan:2015hwa}, muon~\cite{Sirunyan:2018fpa}, charged hadron, neutral hadron, and photon candidates.
These particles are defined with respect to the primary \pp interaction vertex, which is the vertex with the largest value of summed physics-object $\pt^2$.
The physics objects are the jets, clustered using the anti-\kt algorithm~\cite{Cacciari:2008gp,Cacciari:2011ma} with the tracks assigned to candidate vertices as inputs, and the associated missing transverse momentum, taken as the negative vector sum of the \pt of those jets.
Charged-hadron candidates not originating from the selected primary vertex in the event are discarded from the list of reconstructed particles.

Electron candidates are reconstructed using tracking and ECAL information, by combining the clusters of energy deposits in the ECAL with charged tracks~\cite{Khachatryan:2015hwa}.
The electron identification is performed using shower shape variables, track-cluster matching variables, and track quality variables.
The selection is optimized to identify electrons from the decay of \PW and \PZ bosons while rejecting electron candidates originating from jets.
To reject electrons originating from photon conversions inside the detector, electrons are required to have all possible measurements in the innermost tracker layers and to be incompatible with any conversion-like secondary vertices.
Reconstruction of muon candidates is done by geometrically matching tracks from measurements in the muon system and tracker, and fitting them to form a global muon track.
Muons are identified using the quality of the geometrical matching and the quality of the tracks~\cite{Sirunyan:2018fpa}.

In all three running periods, the selected lepton candidates are required to satisfy $\pt > 30 \, \allowbreak (20)$\GeV for the leading (subleading) lepton, and $\abs{\eta} < 2.4$, and to be isolated.
To obtain a measure of isolation for leptons with  $\pt<50\GeV$, a cone with radius $\Delta R=\sqrt{\smash[b]{(\Delta\eta)}^2+\smash[b]{(\Delta\phi)}^2}=0.2$ (where $\phi$ is the azimuthal angle in radians) is constructed around the lepton at the event vertex.
For leptons with $\pt>50\GeV$ the radius is reduced to $\Delta R=\mathrm{max}(0.05, 10\GeV/\pt)$.
A lepton is isolated if the scalar \pt sum of photons and neutral and charged hadrons reconstructed by the PF algorithm within this cone is less than 20\% of the lepton \pt, \ie $I_{\text{rel}}<0.2$.
The contribution of neutral particles from pileup interactions is estimated according to the method described in Ref.~\cite{Khachatryan:2015hwa}, and subtracted from the isolation sum.
The remaining selection criteria applied to electrons, muons, and the reconstruction of jets and \ptmiss are described in Ref.~\cite{Sirunyan:2017leh}.
Jets are clustered from PF candidates using the anti-\kt algorithm with a distance parameter of $R=0.4$, and are required to satisfy $\pt > 30\GeV$, $\abs{\eta} < 2.4$, and quality criteria.
A multivariate \PQb tagging discriminator algorithm, DeepCSV~\cite{Sirunyan:2017ezt}, is used to identify jets arising from \PQb quark hadronization and decay (\PQb jets).
The chosen working point has a mistag rate of approximately~1\% for light-flavor jets and a corresponding \PQb tagging efficiency of approximately 70\%, depending on jet \pt and $\eta$.

Scale factors are applied to simulated events to take into account differences between the observed and simulated lepton reconstruction, identification, and isolation, and \PQb tagging efficiencies.
Typical corrections are less than 1\% per lepton and less than 10\% per \PQb-tagged jet.

\section{Search strategy}
\label{sec:strategy}
We select events containing a pair of leptons with opposite charge.
The invariant mass of the lepton pair $m(\ell\ell)$ is required to be greater than 20\GeV to suppress backgrounds with misidentified or nonprompt leptons from the hadronization of (heavy-flavor) jets in multijet events.
Events with additional leptons with $\pt > 15 \GeV$ and satisfying a looser isolation criterion of $I_{\text{rel}}<0.4$ are rejected.
Events with an SF lepton pair that is consistent with the SM Drell--Yan production are removed by requiring $\abs{m_{\PZ} - m(\ell\ell)} > 15\GeV$, where $m_{\PZ}$ is the mass of the \PZ boson.
To further suppress Drell--Yan and other vector boson backgrounds, we require the number of jets (\Njets) to be at least two and, among them, the number of \PQb-tagged jets (\Nbjets) to be at least one.

\begin{figure}[htbp!]
\centering
\includegraphics[width=0.45\textwidth]{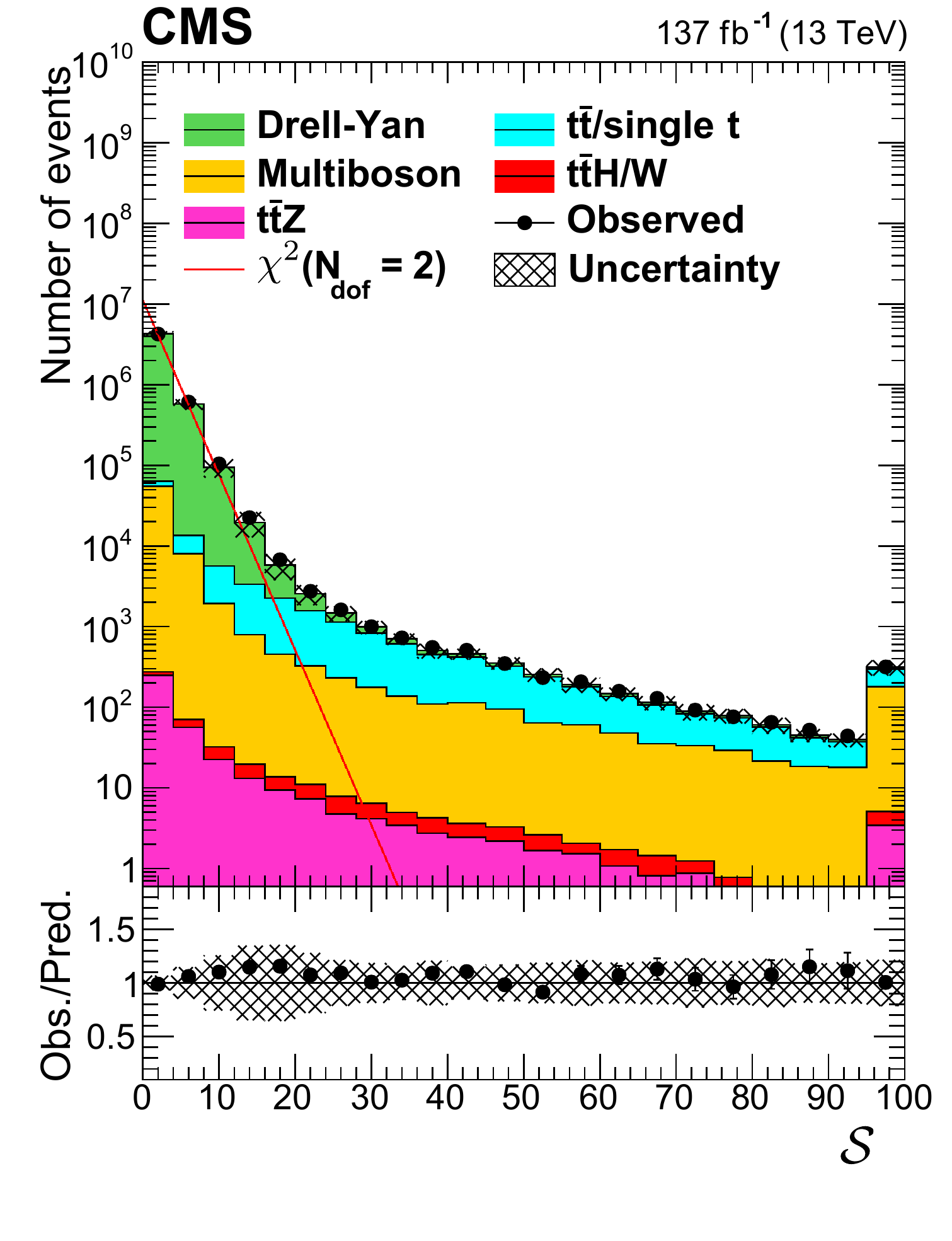}
\caption{
  Distribution of \ptmiss significance \ptmissSig in a $\PZ \to \ell\ell$ selection, requiring an SF lepton pair.
  Points with error bars represent the data, and the stacked histograms the SM backgrounds predicted as described in Section~\ref{sec:backgrounds}, with uncertainty in the SM prediction indicated by the hatched area.
  The red line represents a $\chi^2$ distribution with two degrees of freedom.
  The last bin includes the overflow events.
  The lower panel gives the ratio between the observation and the predicted SM backgrounds.
  The relative uncertainty in the SM background prediction is shown as a hatched band.
  }
\label{fig:METSig}
\end{figure}

We use the \ptmiss significance, denoted as \ptmissSig, to suppress events where detector effects and misreconstruction of particles from pileup interactions are the main source of reconstructed \ptmiss.
In short, the \ptmissSig observable offers an event-by-event assessment of the likelihood that the
observed \ptmiss is consistent with zero.
Using a Gaussian parametrization of the resolutions of the reconstructed objects in the event,
the \ptmissSig observable follows a $\chi^2$-distribution with two degrees of freedom for events with no genuine \ptmiss~\cite{Chatrchyan:2011tn,Khachatryan:2014gga,Sirunyan:2019kia}.
Figure~\ref{fig:METSig} shows the distribution of \ptmissSig in a $\PZ \to \ell\ell$ sample, requiring events with two SF leptons with $\abs{m_{\PZ} - m(\ell\ell)} < 15\GeV$, $\Njets\geq2$ and $\Nbjets=0$.
Events with no genuine \ptmiss, such as from the Drell--Yan process, follow a $\chi^2$ distribution with two degrees of freedom.
Processes with true \ptmiss such as \ttbar or production of two or more \PW or \PZ bosons populate high values of the \ptmissSig distribution.
The algorithm is described in Ref.~\cite{Sirunyan:2019kia} and provides stability of event selection efficiency as a function of the pileup rate.
We exploit this property by requiring $\ptmissSig>12$ in order to suppress the otherwise overwhelming Drell--Yan background in the SF channel.
We further reduce this background by placing a requirement on the azimuthal angular separation of \ptvecmiss and the momentum of the leading~(subleading) jet of $\cos\Delta\phi(\ptmiss, \mathrm{j})<0.80~(0.96)$.
These criteria reject a small background of Drell--Yan events with significantly mismeasured jets.

The event preselection is summarized in Table~\ref{Tab:baselineSel}.
The resulting event sample is dominated by events with top quark pairs that decay to the dilepton final state.
\begin{table}[htb]
\centering
\topcaption{Overview of the event preselection requirements.}
\label{Tab:baselineSel}
\begin{tabular}{ll}
Quantity & Requirement\\\hline
$N_\mathrm{leptons}$ & $=$ 2 (\Pe{} or $\mu$), oppositely charged  \\
$m(\ell\ell)$ & $>$ 20\GeV\\
$\abs{m_{\PZ} - m(\ell\ell)}$ & $>$ 15\GeV, SF only\\
$\Njets$ & $\geq$ 2\\
$\Nbjets$ & $\geq$ 1\\
$\ptmissSig$ & $>$ 12  \\
$\cos\Delta\phi(\ptmiss, \mathrm{j}_{1})$ & $<$ 0.80 \\
$\cos\Delta\phi(\ptmiss, \mathrm{j}_{2})$ & $<$ 0.96 \\
\end{tabular}
\end{table}

The main search variable in this analysis is~\cite{Lester:1999tx,Barr:2003rg}
\ifthenelse{\boolean{cms@external}}{
 \begin{multline}
\mtll = \min_{\ptvecmiss{}^{1} + \ptvecmiss{}^{2} = \ptvecmiss} \Bigl( \max \bigl[ \mt(\ptvec^{\text{vis}1},\\
\ptvecmiss{}^{1}), \mt(\ptvec^{\text{vis}2},\ptvecmiss{}^{2}) \bigr] \Bigr),
\label{eq:MT2def}
\end{multline}
}{
 \begin{equation}
\mtll = \min_{\ptvecmiss{}^{1} + \ptvecmiss{}^{2} = \ptvecmiss} \left( \max \left[ \mt(\ptvec^{\text{vis}1},\ptvecmiss{}^{1}), \mt(\ptvec^{\text{vis}2},\ptvecmiss{}^{2}) \right] \right),
\label{eq:MT2def}
\end{equation}
}
where the choice $\ptvec^{\text{vis}1,2}=\ptvec^{\ell1,2}$ corresponds to the definition introduced in Ref.~\cite{Burns:2008va}.
The alternative choice $\ptvec^{\text{vis}1,2}=\ptvec^{\ell1,2}+\ptvec^{\PQb1,2}$ involves the \PQb-tagged jets and defines \mtlblb.
If only one \PQb-tagged jet is found in the event, the jet with the highest \pt that does not pass the \PQb~tagging selection is taken instead.
The calculation of \mtll and \mtlblb is performed through the algorithm discussed in Ref.~\cite{Lester:2014yga}, assuming vanishing mass for the undetected particles, and follows the description in Ref.~\cite{Sirunyan:2017leh}.
The key feature of the \mtll observable is that it retains a kinematic endpoint at the \PW boson mass for background events from the leptonic decays of two \PW bosons, produced directly or through top quark decay.
Similarly, the \mtlblb observable is bound by the top quark mass if the leptons, neutrinos and \PQb-tagged jets originate from the decay of top quarks.
In turn, signal events from the processes depicted in Fig.~\ref{fig:fey} do not respect the endpoint and are expected to populate the tails of these distributions.

Signal regions based on \mtll, \mtlblb and \ptmissSig are defined to enhance sensitivity to different signal scenarios, and are listed in Table~\ref{table:srdefinition}.
The regions are further divided into different categories based on SF or DF lepton pairs, accounting for the different SM background composition.
The signal regions are defined so that there is no overlap between them, nor with the background-enriched control regions.
\begin{table*}[htb]
\centering
\topcaption{Definition of the signal regions. The regions are further split into SF and DF regions. The preselection in Table~\ref{Tab:baselineSel} is applied to all regions.}
\label{table:srdefinition}
\cmsTable{
\begin{tabular}{ccccc}
 $\mtlblb$ (\GeVns)       & $\ptmissSig$  & $100 < \mtll < 140\GeV$ & $140 < \mtll < 240\GeV$ & $\mtll > 240\GeV$    	                        \\ \hline
\multirow{2}{*}{0--100}   & 12--50        &          SR0            &     SR6                 & \multicolumn{1}{c}{\multirow{6}{*}{SR12} }		\\
                          & $>$50         &          SR1            &     SR7                 & \multicolumn{1}{c}{}{}              					\\
\multirow{2}{*}{100--200} & 12--50        &          SR2            &     SR8                 & \multicolumn{1}{c}{}{}                       \\
                          & $>$50         &          SR3            &     SR9                 & \multicolumn{1}{c}{}{}              					\\
\multirow{2}{*}{ $>$200 } & 12--50        &          SR4            &     SR10                & \multicolumn{1}{c}{}{}              					\\
                          & $>$50         &          SR5            &     SR11                & \multicolumn{1}{c}{}{}              					\\
\end{tabular}
}
\end{table*}

\section{Background predictions}
\label{sec:backgrounds}
Events with an opposite-charge lepton pair are abundantly produced by Drell--Yan and \ttbar processes.
The event selection discussed in Section~\ref{sec:objects} efficiently rejects the vast majority of Drell--Yan events.
Therefore, the major backgrounds from SM processes in the search regions are \PQt/\ttbar events that pass the \mtll threshold because of severely mismeasured \ptmiss or a misidentified lepton.
In signal regions with large \mtll and \ptmissSig requirements, \ttZ events with $\PZ\to\cPgn\cPagn$ are the main SM background.
Remaining Drell--Yan events with large \ptmiss from mismeasurement, multiboson production and other \ttbar/single~\PQt processes in association with a \PW, a \PZ or a Higgs boson (\ttW, \tqZ or \ttH) are sources of smaller contributions.
The background estimation procedures and their corresponding control regions, listed in Table~\ref{table:crdefinition}, are discussed in the following.
\begin{table*}[tph]
  \centering
  \renewcommand*{\arraystretch}{1.3}
  \topcaption{Definition of the control regions. The preselection in Table~\ref{Tab:baselineSel} is applied to all regions.}
  \label{table:crdefinition}
  \begin{tabular}{llll}
    Name    & Definition \\ \hline
    TTCRSF  & \multicolumn{2}{l}{$\mtll<100\GeV$, SF leptons, $\abs{m(\ell\ell)-m_{\PZ}}>15\GeV$}  \\
    TTCRDF  & \multicolumn{2}{l}{$\mtll<100\GeV$, DF leptons} \\ [2\cmsTabSkip]
    TTZ2j2b &                                                      & $\Njets = 2$, $\Nbjets\geq 2$    \\
    TTZ3j1b & $N_{\ell}=3$, $\ptmissSig\geq0$, $\geq 1$ SF lepton pair  & $\Njets = 3$, $\Nbjets = 1$     \\
    TTZ3j2b & with $\abs{m(\ell\ell)-m_{\PZ}}<10\GeV$                  & $\Njets = 3$, $\Nbjets \geq 2$  \\
    TTZ4j1b &                                                      & $\Njets \geq 4$, $\Nbjets = 1$   \\
    TTZ4j2b &                                                      & $\Njets \geq 4$, $\Nbjets\geq 2$ \\ [2\cmsTabSkip]
    \multirow{2}{*}{CR0-CR12}& \multicolumn{2}{l}{Same as SR0-SR12 in Table~\ref{table:srdefinition} but requiring SF leptons, $\abs{m(\ell\ell)-m_{\PZ}}< 15\GeV$,} \\
                             & \multicolumn{2}{l}{$\Nbjets = 0$, and without the $\cos\Delta\phi(\ptmiss, \mathrm{j})$ requirements given in Table~\ref{Tab:baselineSel}.}\\
  \end{tabular}
\end{table*}

\subsection{Top quark background}
Events from the \ttbar process are contained in the $\mtll<100\GeV$ region, as long as the jets and leptons in each event are  identified and their momenta are precisely measured.
Three main sources are identified that promote \ttbar events into the tail of the \mtll distribution.
Firstly, the jet momentum resolution is approximately Gaussian~\cite{Khachatryan:2016kdb} and jet mismeasurements propagate to \ptmiss, which subsequently leads to values of \mtll and \mtlblb that do not obey the endpoint at the mother particle mass.
For events with $\mtll\leq140\GeV$, this \ttbar component is dominant, while it amounts to less than 10\% for signal regions with $\mtll>140\GeV$.
Secondly, significant mismeasurements of the momentum of jets can be caused by the loss of photons and neutral hadrons showering in masked channels of the calorimeters, or neutrinos with high \pt within jets.
For $\mtll>140\GeV$, up to 50\% of the top quark background falls into this category.
The predicted rate and kinematic modeling of these rare non-Gaussian effects in simulation are checked in a control region requiring SF leptons satisfying $\abs{m(\ell\ell)-m_{\PZ}}<15\GeV$.
A 30\% uncertainty covers differences in the tails of the \ptmiss distribution observed in this control region.

Finally, an electron or a muon may fail the identification requirements, or the event may have a $\tau$ lepton produced in a \PW boson decay.
If there is a nonprompt lepton from the hadronization of a bottom quark or a charged hadron misidentified as a lepton selected in the same event, the reconstructed value for \mtll is not bound by the W mass.
To validate the modeling of this contribution, we select events with one additional lepton satisfying loose isolation requirements on top of the selection in Table~\ref{Tab:baselineSel}.
In order to mimic the lost prompt-lepton background, we recompute \mtll by combining each of the isolated leptons with the extra lepton in both the observed and simulated samples.
Since the transverse momentum balance is not significantly changed by the lepton misidentification, the \ptmiss and \metSig observables are not modified.
Events with misidentified electrons or muons from this category constitute up to 40\% of the top quark background prediction for $\mtll>140\GeV$.
We see good agreement between the observed and simulated kinematic distributions, indicating that the simulation describes such backgrounds well.
Based on the statistical precision in the highest \mtll regions, we assign a 50\% uncertainty to this contribution.

The \ttbar normalization is measured \textit{in situ} by including a signal-depleted control region defined by $\mtll<100\GeV$ in the signal extraction fit, yielding a scale factor for the \ttbar prediction of $1.02 \pm 0.04$.
The region is split into DF~(TTCRDF) and SF channels~(TTCRSF).
Events with a \PZ boson candidate are rejected in the latter.

\subsection{Top quark + X background}
Top quarks produced in association with a boson (\ttZ, \ttW, \ttH, \tqZ) form an irreducible background, if the boson decays to leptons or neutrinos.
The $\PZ\to \cPgn\cPagn$ decay in the \ttZ process provides genuine \ptmiss and is the dominant background component at high values of \mtll.
The decay mode
$
\ttZ \to  (\cPqb\ell^{\pm}\cPgn)(\PQb\text{jj})(\ell^{\pm}\ell ^{\mp})
$
is used to measure the normalization of this contribution.
The leading, subleading, and trailing lepton \pt are required to satisfy thresholds of $40$, $20$, and $20\GeV$, respectively.
The invariant mass of two SF leptons with opposite charge is required to satisfy the tightened requirement $\abs{m(\ell\ell) - m_{\PZ}} <10\GeV$.
The shape of the distribution of $\pt(\PZ)$ has recently been measured in the 2016 and 2017 data sets~\cite{CMS:2019too} and is well described by simulation.
Five control regions requiring different \Njets and \Nbjets combinations are defined in Table~\ref{table:crdefinition} and labeled TTZ2j2b--TTZ4j2b.
They are included in the signal extraction fit, in which the simulated number of \ttZ events is found to be scaled up by a factor of $1.22 \pm 0.25$, consistent with the initial prediction.

\subsection{Drell--Yan and multiboson backgrounds}
In order to measure the small residual Drell--Yan contribution that passes the event selection, we select dilepton events according to the criteria listed in Table~\ref{Tab:baselineSel} except that we invert the \PZ boson veto, the \PQb jet requirements, and remove the angular separation requirements on jets and \ptvecmiss.
We expect from the simulation that the selection is dominated by the Drell--Yan and multiboson events.
For each SF signal region, we define a corresponding control region with the selections above and the signal region requirements on \mtll, \mtlblb, and \ptmissSig.
The regions are labeled CR0--CR12 in Table~\ref{table:crdefinition} and are included in the signal extraction fit.
The \mtlblb observable  is calculated in these regions using the two highest \pt jets.
The scale factors for the Drell--Yan and multiboson background components are found to be $1.18 \pm 0.28$ and $1.35 \pm 0.32$, respectively.

\begin{figure*}[tph]
  \centering
  \includegraphics[width = .32\linewidth]{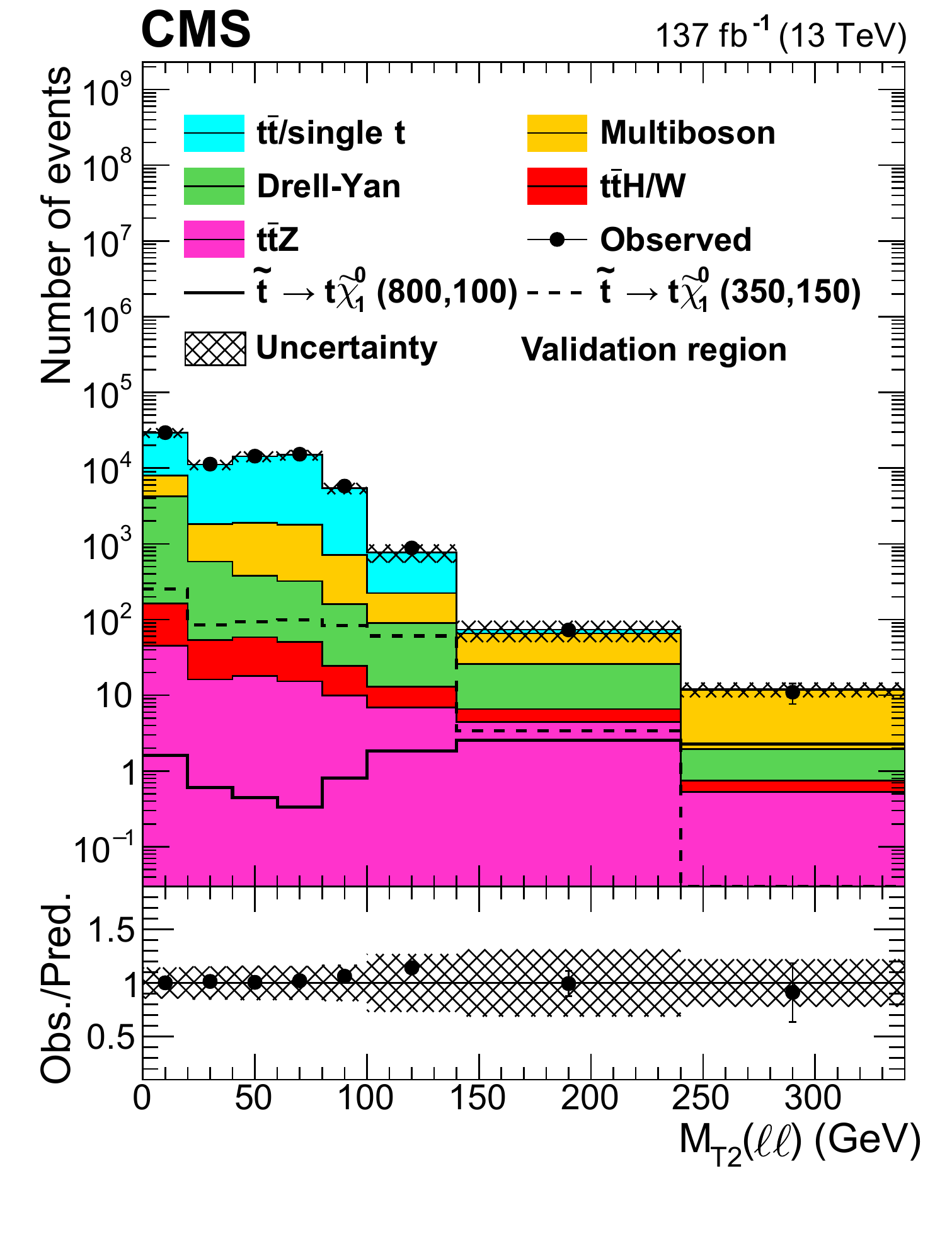}
  \includegraphics[width = .32\linewidth]{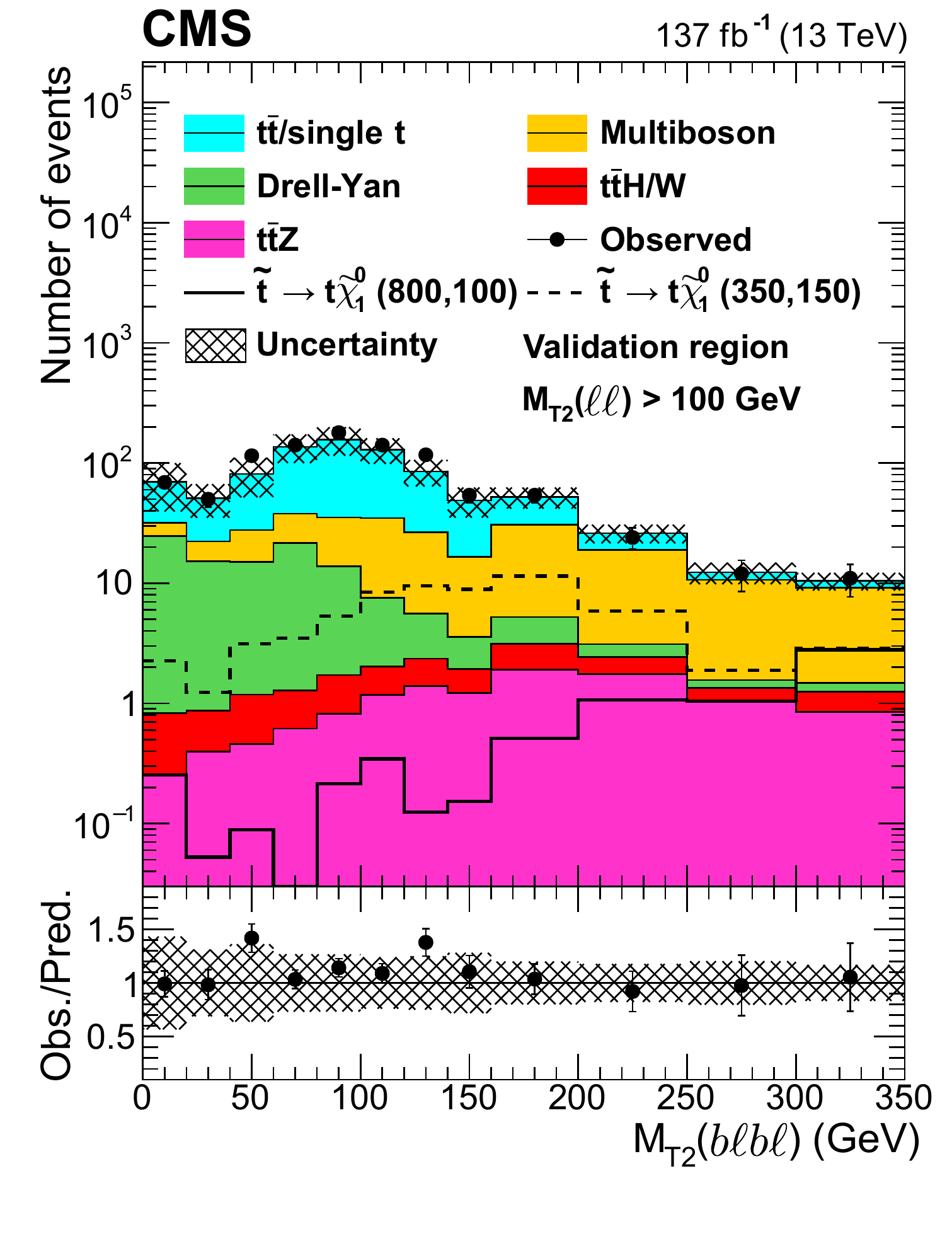}
  \includegraphics[width = .32\linewidth]{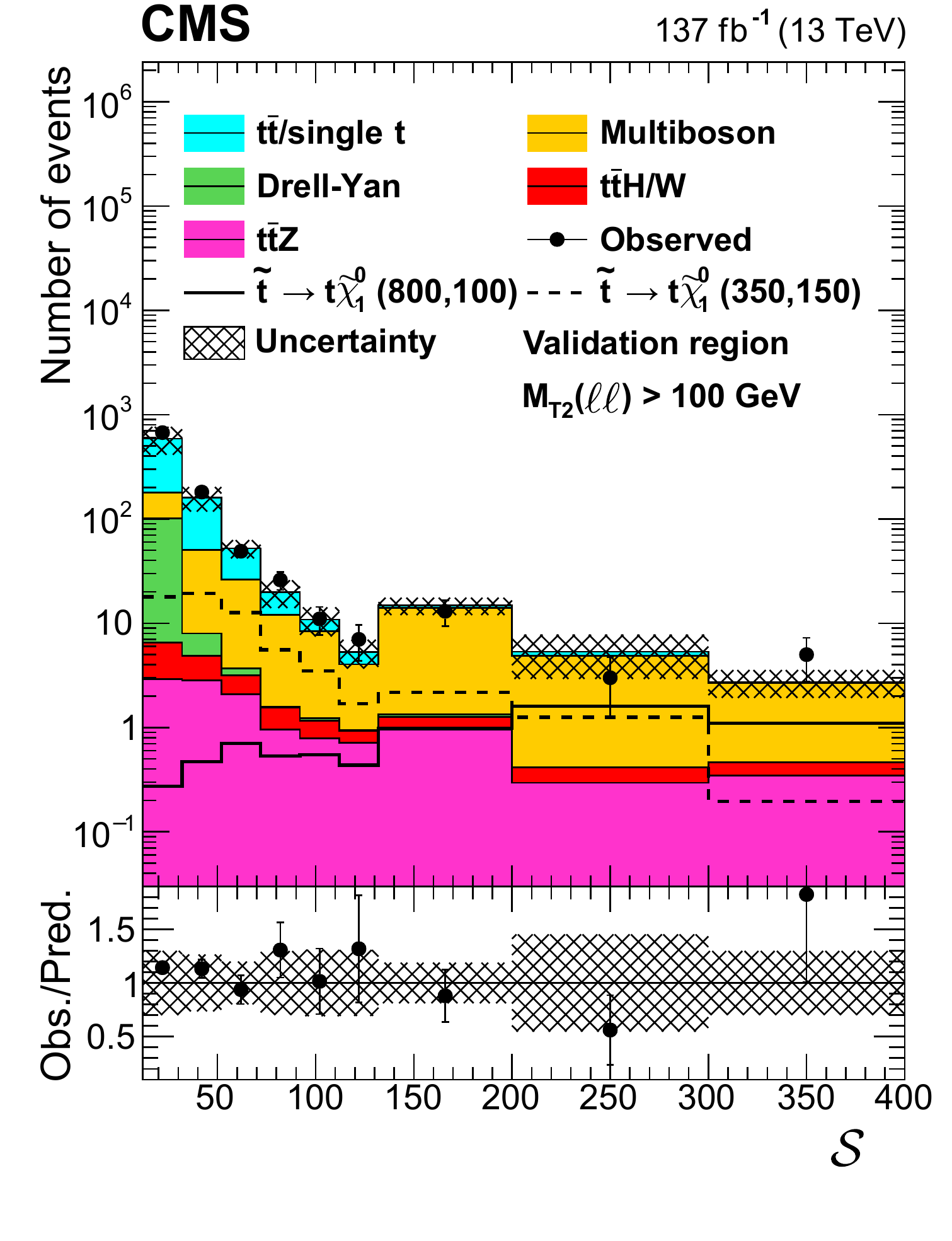}
  \caption{ The \mtll, \mtlblb, and \metSig distributions in the validation regions requiring $\Njets\geq 2$ and $\Nbjets=0$, combining the SF and DF channels.
  All other event selection requirements are applied.
  For the \mtlblb and \metSig distributions, $\mtll > 100\GeV$ is required.
  The individual processes are scaled using their measured respective scale factors, as described in the text.
  The hatched band represents the experimental systematic uncertainties and the uncertainties in the scale factors.
  The last bin in each distribution includes the overflow events.
  The lower panel gives the ratio between the observation and the predicted SM backgrounds.
  The relative uncertainty in the SM background prediction is shown as a hatched band.
    }
  \label{fig:CR-2j0bS12-comb}
\end{figure*}

The good modeling of the multiboson and \ttbar processes, including potential sources of anomalous \ptmiss, is demonstrated in a validation region requiring $\Njets\geq 2$ and $\Nbjets=0$ and combining the SF and DF channels.
The observed distributions of the search variables are compared with the simulated distributions in Fig.~\ref{fig:CR-2j0bS12-comb}.
The hatched band includes the experimental systematic uncertainties and the uncertainties in the background normalizations.

\section{Systematic uncertainties}
\label{sec:systematics}
Several experimental uncertainties affect the signal and background yield estimations.
The efficiency of the trigger selection ranges from 95 to 99\% with uncertainties lower than 2.3\% in all signal and control regions.
Offline lepton reconstruction and selection efficiencies are measured using $\PZ\to \ell\ell$ events in bins of lepton \pt and $\eta$.
These measurements are performed separately in the observed and simulated data sets, with efficiency values ranging from 70 to 80\%.
Scale factors are used to correct the efficiencies measured in simulated events to those in the observed data.
The uncertainties in these scale factors are less than 3\% per lepton and less than 5\% in most of the search and control regions.

Uncertainties in the event yields resulting from the calibration of the jet energy scale are estimated by shifting the jet momenta in the
simulation up and down by one standard deviation of the jet energy corrections. Depending on the jet \pt and $\eta$,
the resulting uncertainty in the simulated yields from the jet energy scale is typically 4\%, except in the lowest regions in \mtll close to the $m_\PW$ threshold where it can be as high as 20\%.
In addition, the energy scale of deposits from soft particles that are not clustered in jets are varied within their uncertainties, and the resulting uncertainty reaches 7\%.
The \PQb tagging efficiency in the simulation is corrected using scale factors determined from the observed data~\cite{Sirunyan:2017ezt}, and uncertainties are propagated to all simulated events.
These contribute an uncertainty of up to 7\% in the predicted yields, depending on the \pt, $\eta$ and origin of the \PQb-tagged jet.

The effect of all the experimental uncertainties described above is evaluated for each of the simulated processes in all signal regions, and is considered correlated across the analysis bins and simulated processes.

The uncertainties in the normalizations of the single top and \ttbar, \ttZ, Drell--Yan, and multiboson backgrounds are discussed in Section~\ref{sec:backgrounds}.
Finally, the uncertainty in the integrated luminosity is 2.3--2.5\%~\cite{lumipas2016,lumipas2017,lumipas2018}.

Additional systematic uncertainties affect the modeling in simulation of the various processes, discussed in the following.
All simulated samples are reweighted according to the distribution of the true number of interactions at each bunch crossing.
The uncertainty in the total inelastic \pp cross section leads to uncertainties of 5\% in the expected yields.

For the \ttbar and \ttZ backgrounds, we determine the event yield changes resulting from varying the renormalization scale~($\mu_\mathrm{R}$) and the factorization scale~($\mu_\mathrm{F}$) up and down by a factor of two, while keeping the overall normalization constant.
The combinations of variations in opposite directions are disregarded.
We assign as the uncertainty the envelope of the considered yield variations, treated as uncorrelated among the background processes.
Uncertainties in the PDFs can have a further effect on the simulated \mtll shape.
We determine the change of acceptance in the signal regions using the PDF variations and assign the envelope of these variations---less than 4\%---as a correlated uncertainty~\cite{Butterworth:2015oua}.

The contributions to the total uncertainty in the estimated backgrounds are summarized in Table~\ref{tab:post-fit}, which provides the maximum uncertainties over all signal regions
and the typical values, defined as the 90\% quantile of the uncertainty values in all signal regions.
\begin{table*}
  \centering
  \topcaption{Typical values (90\% quantiles) and maximum values of the systematic uncertainties in all signal regions.
       }
  \label{tab:post-fit}
  \begin{tabular}{lcc}
  Systematic uncertainty & Typical (\%) & Max (\%) \\\hline
  Integrated luminosity               & $ 2      $ & $ 2      $   \\
  Pileup modeling                     & $ 5      $ & $ 7      $   \\
  Jet energy scale                    & $ 4      $ & $ 20     $   \\
  Jet energy resolution               & $ 3      $ & $ 4      $   \\
  \cPqb tagging efficiency            & $ 2      $ & $ 3      $   \\
  \cPqb tagging mistag rate           & $ 1      $ & $ 7      $   \\
  Trigger efficiency                  & $ 1      $ & $ 2      $   \\
  Lepton identification efficiency    & $ 3      $ & $ 5      $   \\
  Modeling of unclustered energy      & $ 3      $ & $ 7      $   \\
  Non-Gaussian jet mismeasurements    & $ 6      $ & $ 6      $   \\
  Misidentified or nonprompt leptons     & $ 5      $ & $ 5      $   \\
  \ttbar normalization                & $ 9      $ & $ 9      $   \\
  \ttZ normalization                  & $ 10     $ & $ 14     $   \\
  Multiboson background normalization & $ 4      $ & $ 8      $   \\
  \ttH/\PW background normalization       & $ 5      $ & $ 8      $   \\
  Drell--Yan normalization            & $ 3      $ & $ 8      $   \\
  Parton distribution functions       & $ 2      $ & $ 4      $   \\
  $\mu_\mathrm{R}$ and $\mu_\mathrm{F}$ choice      & $ 7      $ & $ 11     $   \\
\end{tabular}
\end{table*}

For the small contribution from \ttbar production in association with a \PW or a Higgs boson, we take an uncertainty of 20\% in the
cross section based on the variations of the generator scales and the PDFs.

Most of the sources of systematic uncertainty in the background estimates affect the prediction of the signal as well, and these are evaluated separately for each mass configuration of the considered simplified models.
We further estimate the effect of missing higher-order corrections for the signal acceptance by varying $\mu_\mathrm{R}$ and $\mu_\mathrm{F}$~\cite{Catani2003zt,Cacciari2003fi,Kalogeropoulos:2018cke} and find that those uncertainties are below 10\%.
The modeling of initial-state radiation (ISR) is relevant for the SUSY signal simulation in cases where the mass difference between the top squark and the LSP is small.
The ISR reweighting is based on the number of ISR jets ($N_\mathrm{J}^{\text{ISR}}$) so as to make the predicted jet multiplicity distribution agree with that observed.
The comparison is performed in a sample of events requiring two leptons and two \PQb-tagged jets.
The reweighting procedure is applied to SUSY MC events
and factors vary between 0.92 and 0.51 for $N_\mathrm{J}^{\text{ISR}}$ between 1 and 6.
We take one half of the deviation from unity as the systematic uncertainty in these reweighting factors, correlated across search regions.
It is generally found to have a small effect, but can reach 30\% for compressed mass configurations.
An uncertainty from potential differences of the modeling of \ptmiss in the fast simulation of the CMS detector is evaluated by comparing
the reconstructed \ptmiss with the \ptmiss obtained using generator-level information. This uncertainty ranges up to 20\% and only affects the SUSY signal samples.
For these samples, the scale factors and uncertainties for the tagging efficiency of \PQb jets and leptons are evaluated separately.
Typical uncertainties in the scale factors are below 2\% for \PQb-tagged jets, and between 1 and 7\% for leptons.

\section{Results}
\label{sec:results}
\begin{figure*}[tbp]
\centering
\includegraphics[width=.32\textwidth]{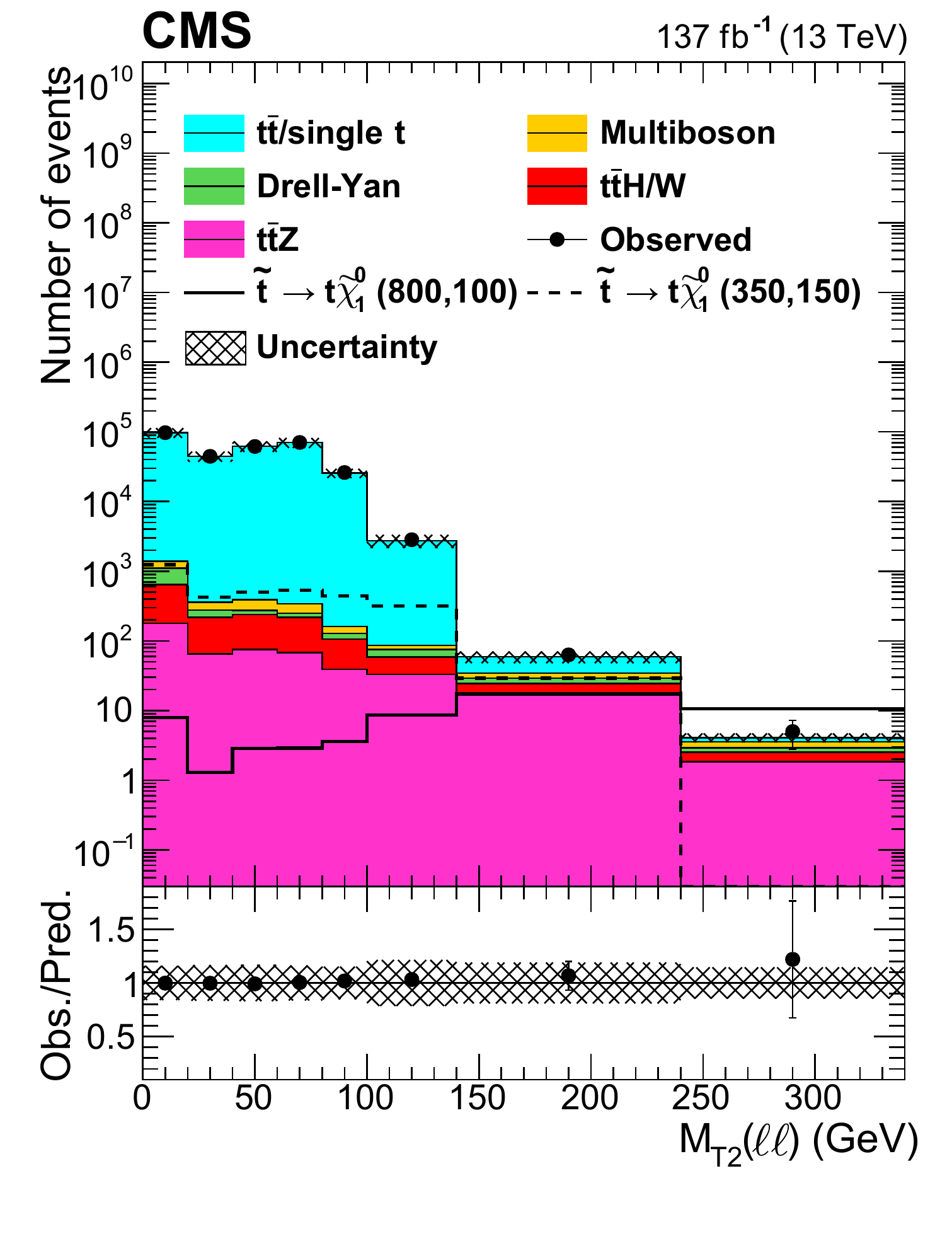}
\includegraphics[width=.32\textwidth]{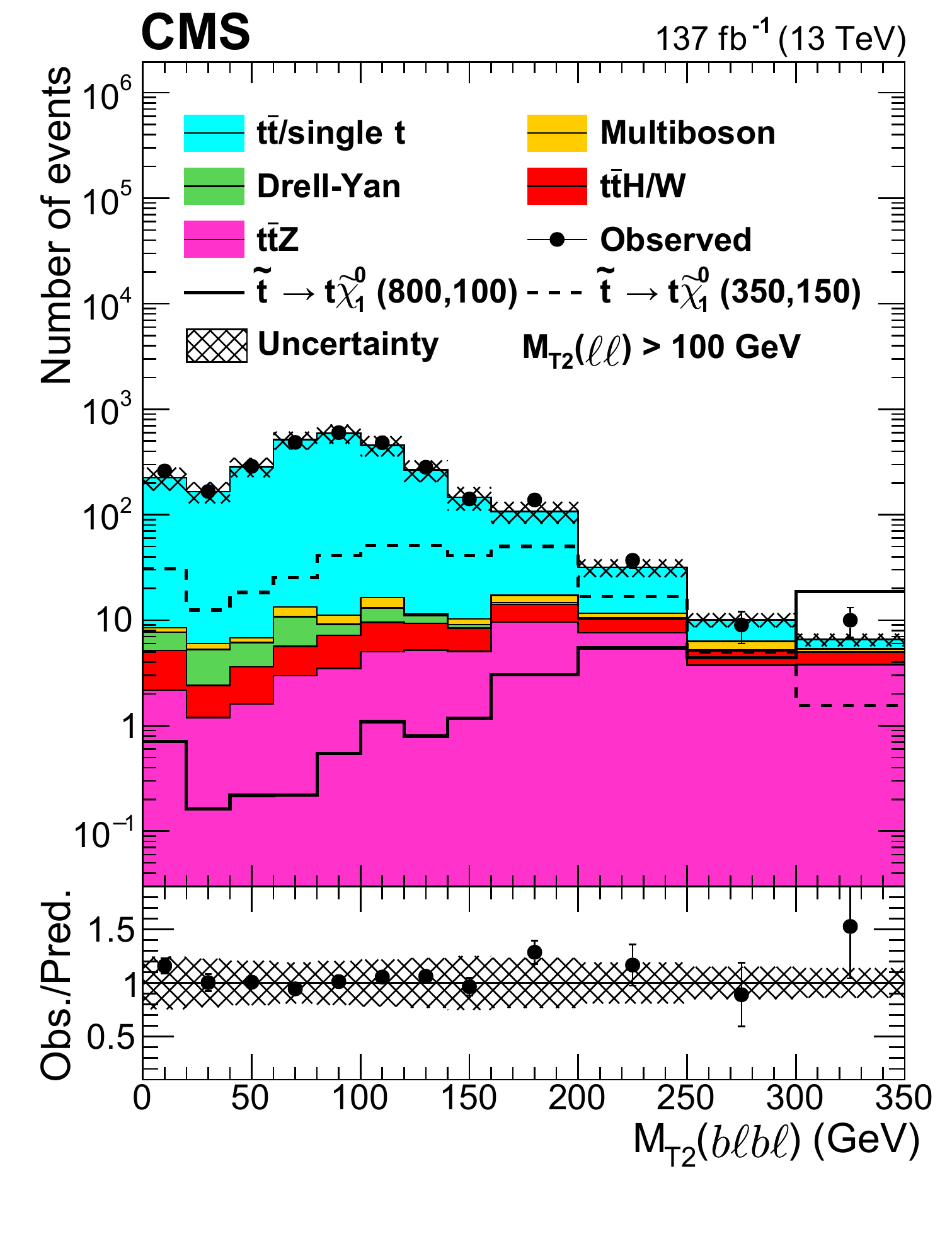}
\includegraphics[width=.32\textwidth]{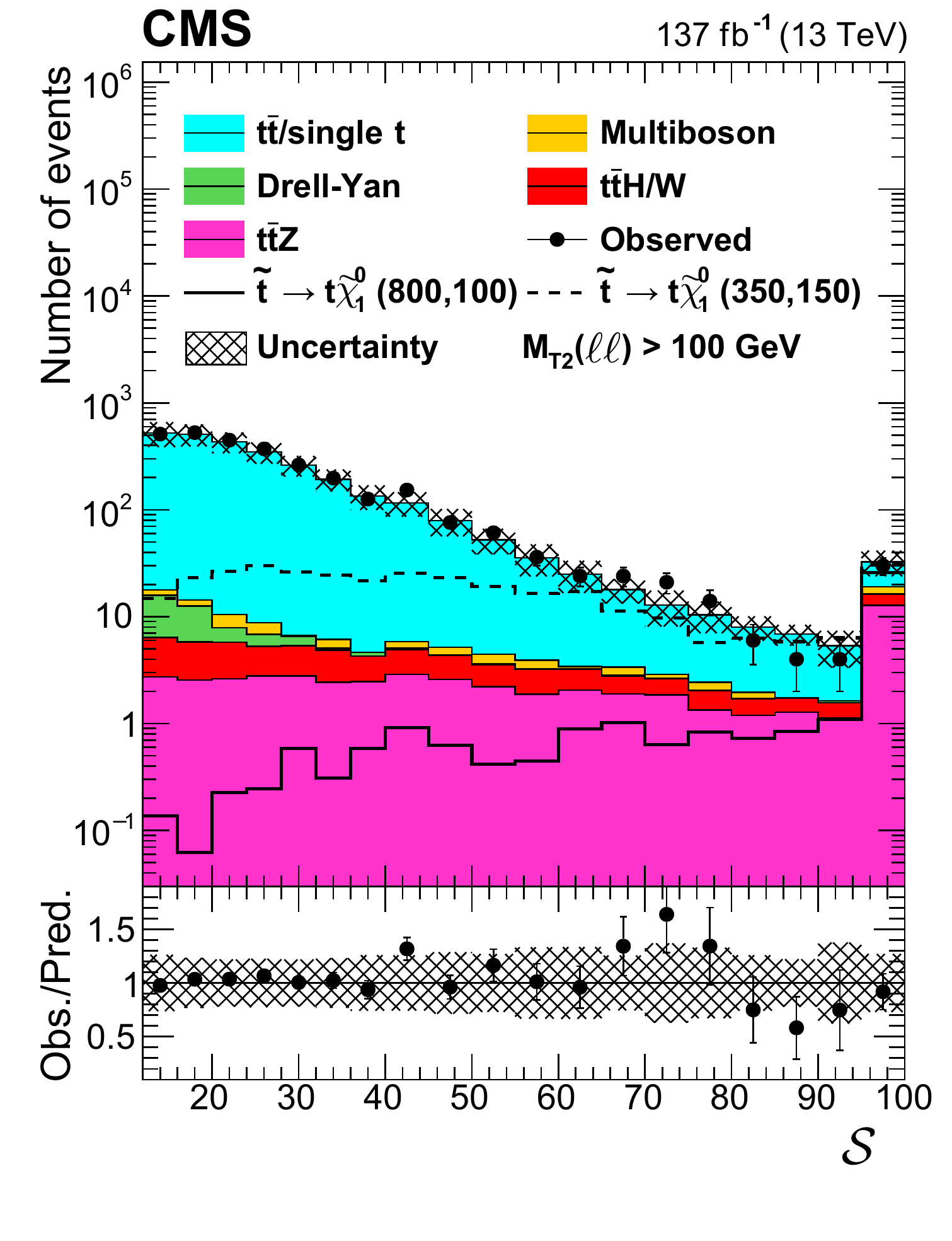}
\caption{
Distributions of \mtll (left), \mtlblb (middle), and \metSig (right) for all lepton flavors for the preselection defined in Table~\ref{Tab:baselineSel}.
Additionally, $\mtll > 100\GeV$ is required for the \mtlblb and \metSig distributions.
The last bin in each distribution includes the overflow events.
The lower panel gives the ratio between the observation and the predicted SM backgrounds and the relative uncertainty in the SM background prediction is shown as a hatched band.
}
\label{fig:SR-2j1bS12-comb}
\end{figure*}

\begin{figure*}[htbp!]
\centering
\includegraphics[width=0.98\textwidth]{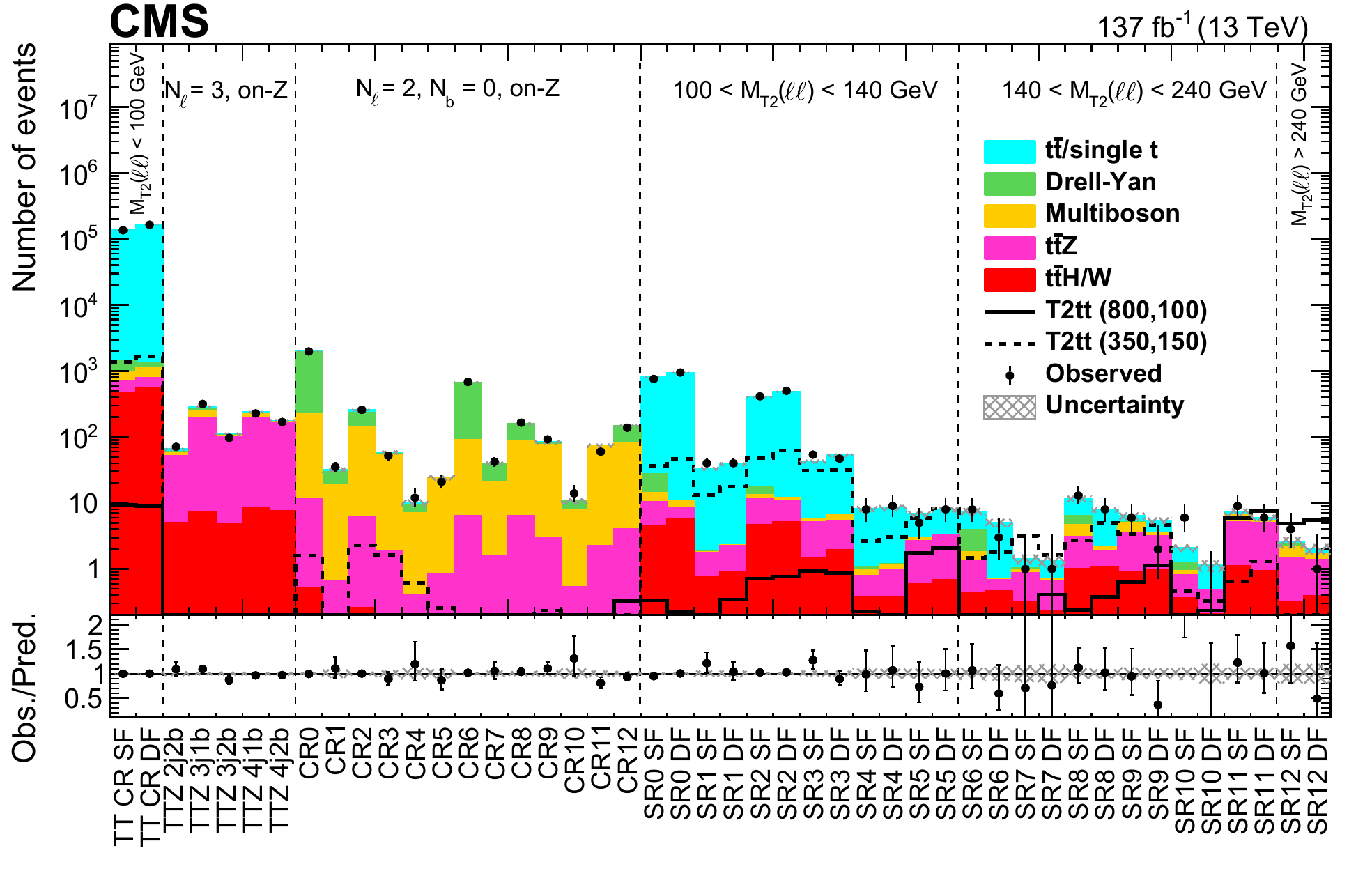}
\caption{
  Predicted and observed yields in the signal and control regions as defined in Tables~\ref{table:srdefinition} and \ref{table:crdefinition}.
  The control regions TTCRSF and TTCRDF are defined by $\mtll < 100\GeV$ and are used to constrain the \ttbar normalization.
  The \ttZ control regions employ a 3 lepton requirement in different \Njets and \Nbjets bins.
  The dilepton invariant mass and \Nbjets selections are inverted for CR0--CR12 in order to constrain the Drell--Yan and multiboson normalizations, using only the SF channel.
  The lower panel gives the ratio between the observation and the predicted SM backgrounds.
  The hatched band reflects the post-fit systematic uncertainties.}
\label{fig:overall_results_postfit}
\end{figure*}

Good agreement between the SM-predicted and observed \mtll, \mtlblb, and \metSig distributions is found, as shown in Fig.~\ref{fig:SR-2j1bS12-comb}.
No significant deviation from the SM prediction is observed in any of the signal regions as shown in Fig.~\ref{fig:overall_results_postfit}.
The observed excess events in SR10SF are found to be close to the signal region selection thresholds.
To perform the statistical interpretations, a likelihood function is formed with Poisson probability functions for all data regions.
The control and signal regions as depicted in Fig.~\ref{fig:overall_results_postfit} are included.
The correlations of the uncertainties are taken into account as described in Section~\ref{sec:systematics}.
A profile likelihood ratio in the asymptotic approximation~\cite{Cowan:2010js} is used as the test statistic. Upper limits on the production cross section are calculated at 95\%
confidence level (\CL) according to the asymptotic \CLs criterion~\cite{Junk1999,ClsCite}.

\begin{figure*}[htbp]
\centering
\includegraphics[width=.49\textwidth]{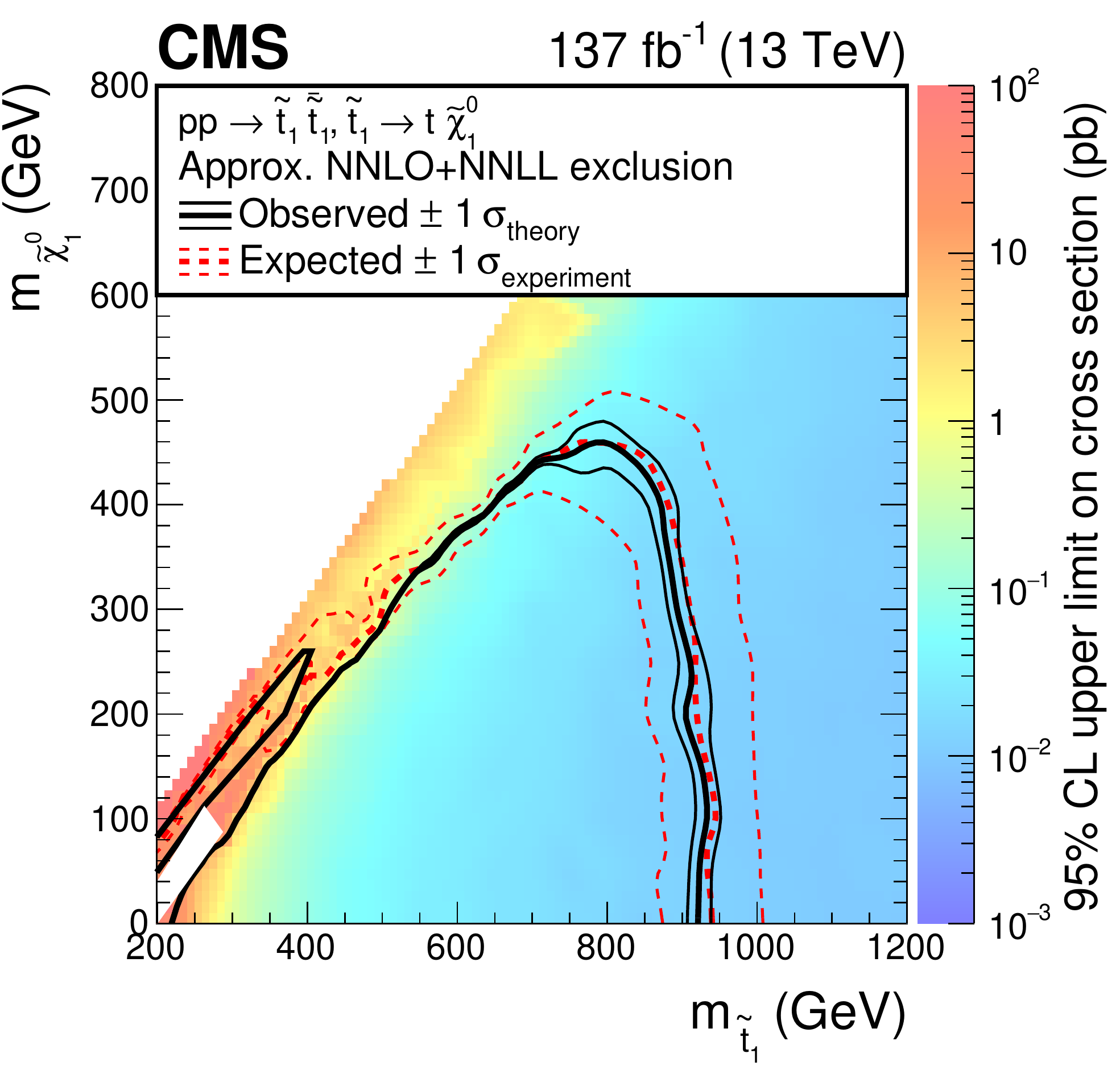}
\includegraphics[width=.49\textwidth]{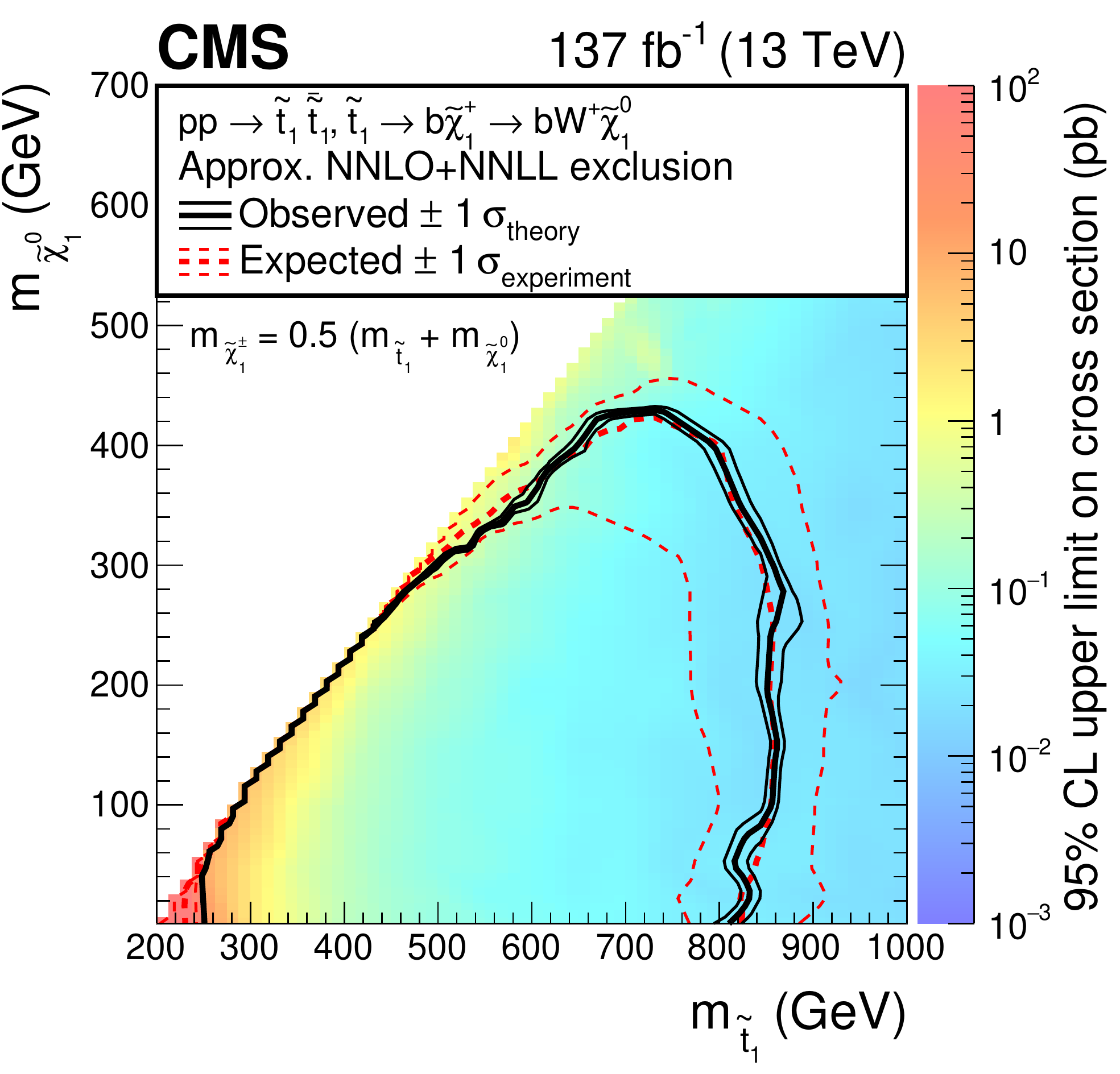}
\caption{Expected and observed limits for the \Ttt model with $\PSQtDo \to \PQt\PSGczDo$ decays (left) and for the \TbW model with $\PSQtDo \to \PQb\PSGcpDo \to \PQb\PWp\PSGczDo$ decays (right) in the $m_{\PSQtDo}$-$m_{\PSGczDo}$ mass plane.
The color indicates the 95\% \CL upper limit on the cross section at each point in the plane.
The area below the thick black curve represents
the observed exclusion region at 95\% \CL assuming 100\% branching fraction for the decays of the SUSY particles,
while the dashed red lines indicate the expected limits at 95\% \CL and the region containing 68\% of the distribution of limits expected under the background-only hypothesis.
The thin black lines show the effect of the theoretical uncertainties in the signal cross section.
The small white area on the diagonal in the left figure corresponds to
configurations where the mass difference between \PSQtDo and \PSGczDo is very close to the top quark mass.
In this region the signal acceptance strongly depends on the \PSGczDo mass and is therefore hard to model.
}
\label{fig:limit_T2tt_T2bW}
\end{figure*}

\begin{figure*}[htb]
\centering
\includegraphics[width=.49\linewidth]{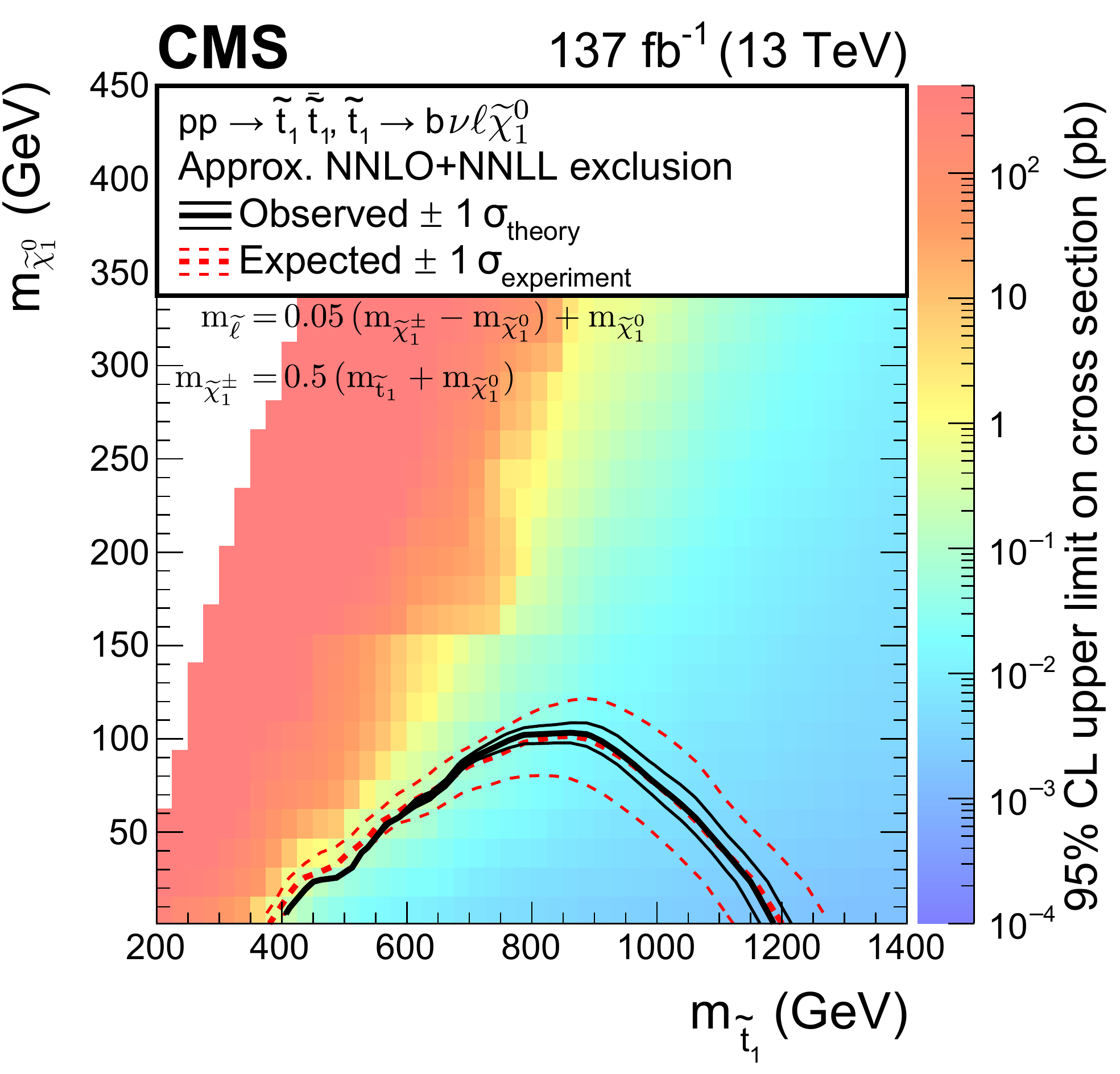}
\includegraphics[width=.49\linewidth]{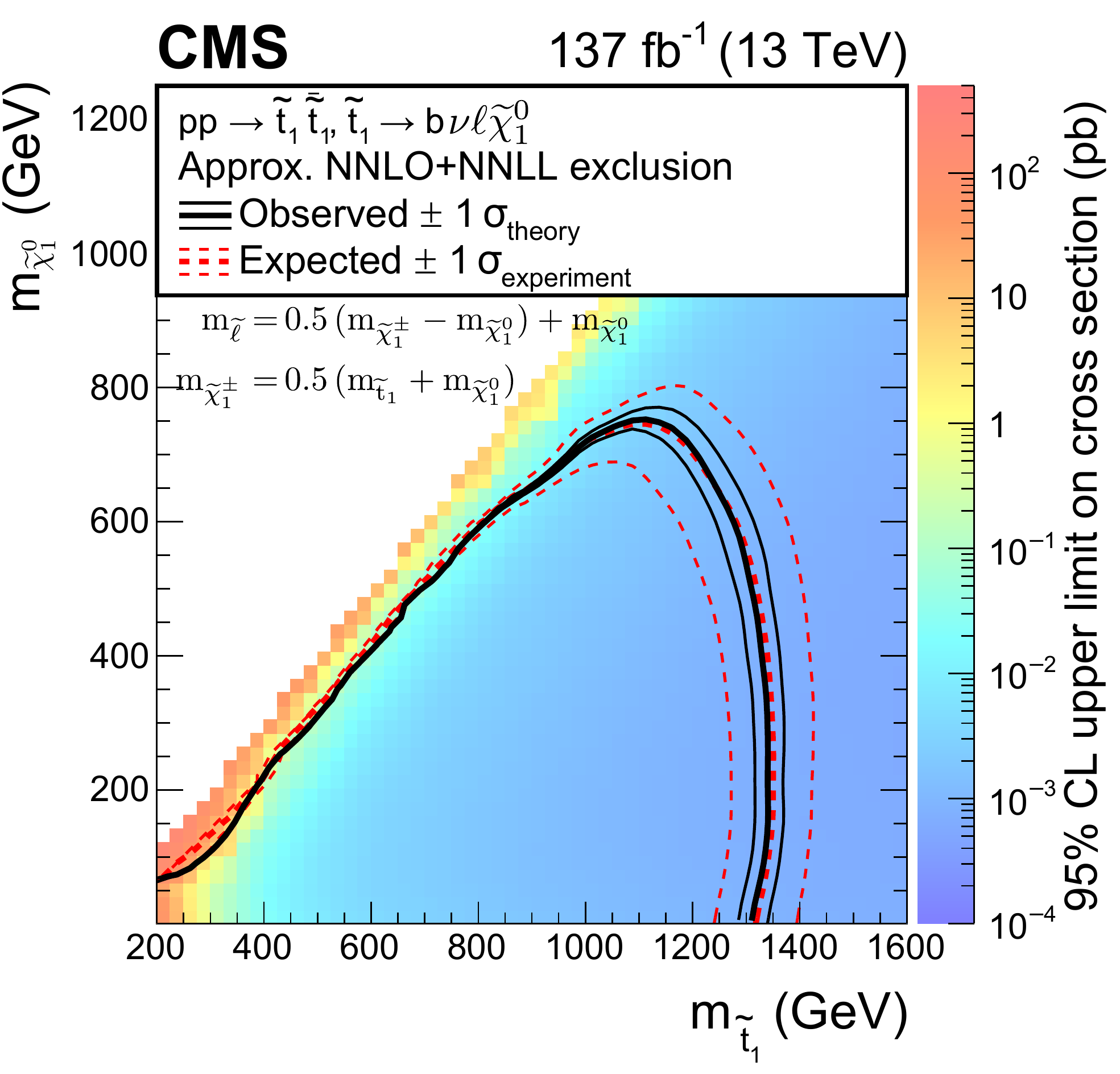}\\
\includegraphics[width=.49\linewidth]{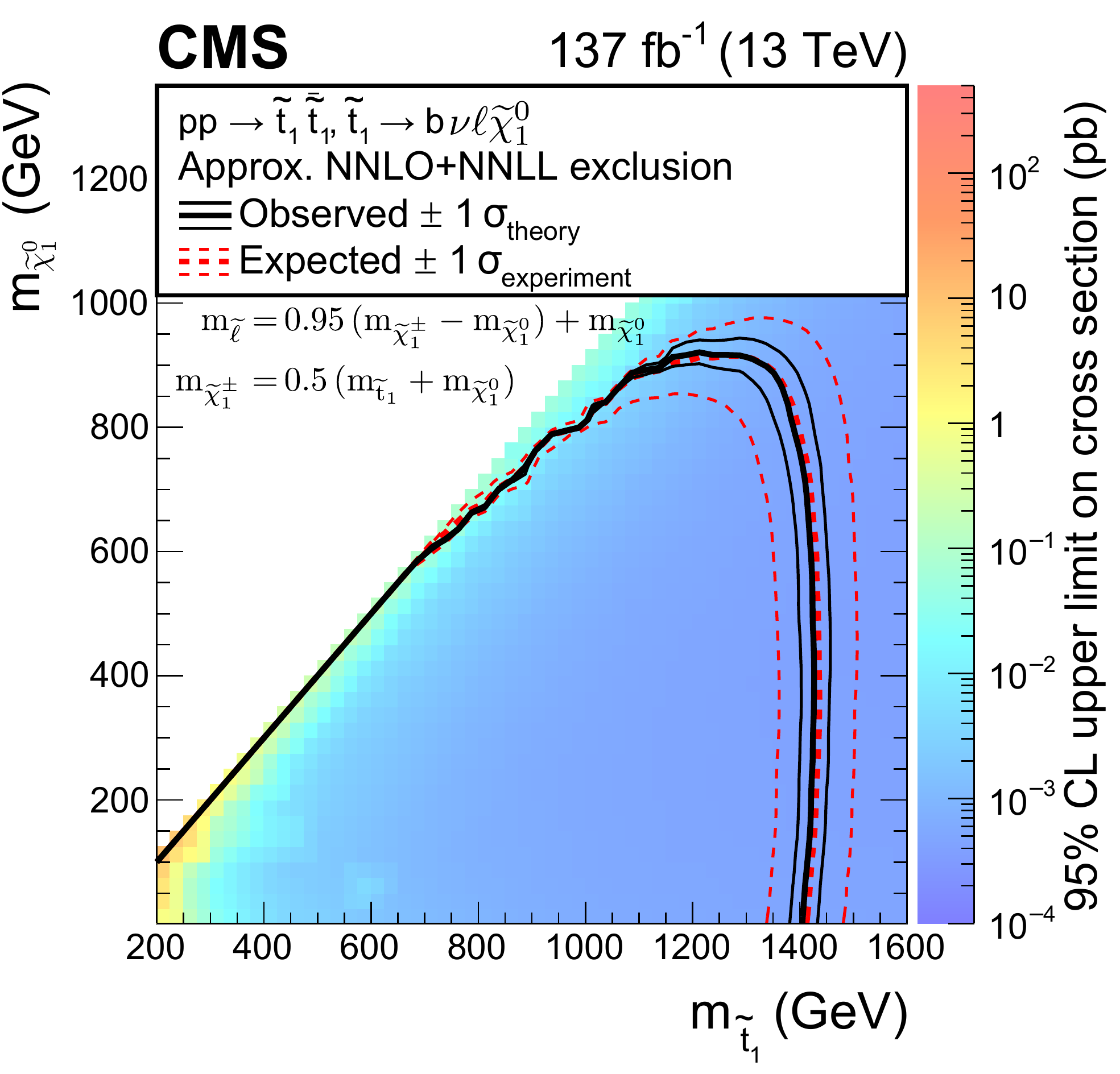}
\caption{Expected and observed limits for the \Tbbllnunu model with $\PSQtDo \to \PQb\PSGcpDo \to \PQb\PGn\widetilde{\ell} \to \PQb\PGn\ell\PSGczDo$ decays in the $m_{\PSQtDo}$-$m_{\PSGczDo}$ mass plane for
three different mass configurations defined by $m_{\widetilde{\ell}} = x \, (m_{\PSGcpDo} - m_{\PSGczDo}) + m_{\PSGczDo}$ with $x=0.05$ (upper left), $x=0.50$ (upper right), and $x=0.95$ (lower).
The description of curves is the same as in the caption of Fig.~\ref{fig:limit_T2tt_T2bW}.
}
\label{fig:limit_T8bbllnunu}
\end{figure*}

The results shown in Fig.~\ref{fig:overall_results_postfit} are interpreted in the context of simplified SUSY models of top squark production followed by a decay to top quarks and neutralinos~(\Ttt), via an intermediate chargino~(\TbW), and via an additional intermediate slepton~($\Tbbllnunu$).
These interpretations are presented on the $m_{\PSQtDo}$-$m_{\PSGczDo}$ plane in Figs.~\ref{fig:limit_T2tt_T2bW} and~\ref{fig:limit_T8bbllnunu}.
The color on the $z$ axis indicates the 95\% \CL upper limit on the cross section at each point in the $m_{\PSQtDo}$-$m_{\PSGczDo}$ plane.
The area below the thick black curve represents the observed exclusion region at 95\% \CL
assuming 100\% branching fraction for the decays of the SUSY particles.
The thick dashed red lines indicate the expected limit at 95\% \CL, while the region containing 68\% of the distribution of limits expected under the background-only hypothesis is bounded by thin dashed red lines.
The thin black lines show the effect of the theoretical uncertainties in the signal cross section.
In the \Ttt model we exclude mass configurations with $m_{\PSGczDo}$ up to 450\GeV and  $m_{\PSQtDo}$ up to 925\GeV, assuming that the top quarks are unpolarized, thus improving by
approximately 125\GeV in $m_{\PSQtDo}$ the results presented on a partial data set in Ref.~\cite{Sirunyan:2017leh}.
The observed upper limit on the top squark cross section improved by approximately 50\% for most mass configurations.
The result for the \TbW model is shown in Fig.~\ref{fig:limit_T2tt_T2bW}~(right) and the results for \Tbbllnunu models are shown in Fig.~\ref{fig:limit_T8bbllnunu}.
We exclude mass configurations with $m_{\PSGczDo}$ up to 420\GeV and  $m_{\PSQtDo}$ up to 850\GeV in the \TbW model, extending the exclusion limits set in Ref.~\cite{Sirunyan:2017leh} by approximately 100\GeV in $m_{\PSQtDo}$.
The sensitivity in the \Tbbllnunu model strongly depends on the intermediate slepton mass and is largest when $x = 0.95$ in $m_{\widetilde{\ell}} = x \, (m_{\PSGcpDo} - m_{\PSGczDo}) + m_{\PSGczDo}$.
In this case, excluded masses reach up to 900\GeV for $m_{\PSGczDo}$ and 1.4\TeV for $m_{\PSQtDo}$.
These upper limits decrease to 750\GeV for $m_{\PSGczDo}$ and 1.3\TeV for $m_{\PSQtDo}$ when $x=0.5$
and to 100\GeV for $m_{\PSGczDo}$ and 1.2\TeV for $m_{\PSQtDo}$ when $x=0.05$.
In this model, the improvement upon previous results from Ref.~\cite{Sirunyan:2017leh} is approximately 100\GeV in $m_{\PSQtDo}$, and up to 100\GeV in $m_{\PSGczDo}$.

\section{Summary}
\label{sec:summary}
A search for top squark pair production in final states with two opposite-charge leptons, \PQb jets, and significant missing transverse momentum (\ptmiss) is presented.
The data set of proton-proton collisions corresponds to an integrated luminosity of \lumiGolden and was collected with the CMS detector at a center-of-mass energy of 13\TeV.
Transverse mass variables and the significance of \ptmiss are used to efficiently suppress backgrounds from standard model processes.
No evidence for a deviation from the expected background is observed.
The results are interpreted in several simplified models for supersymmetric top squark pair production and decay.

In the \Ttt model with $\PSQtDo \to \PQt\PSGczDo$ decays, $\PSQtDo$ masses up to 925\GeV and $\PSGczDo$ masses up to 450\GeV are excluded.
In the \TbW model with $\PSQtDo \to \PQb\PSGcpDo \to \PQb\PWp\PSGczDo$ decays, $\PSQtDo$ masses up to 850\GeV and $\PSGczDo$ masses up to 420\GeV are excluded, assuming the chargino mass to be the mean of the $\PSQtDo$ and $\PSGczDo$ masses.
In the \Tbbllnunu model with decays $\PSQtDo \to \PQb\PSGcpDo \to \PQb\PGn\widetilde{\ell} \to \PQb\PGn\ell\PSGczDo$, therefore 100\% branching fraction to dilepton final states, the sensitivity depends on the intermediate particle masses. With the chargino mass again taken as the mean of the $\PSQtDo$ and $\PSGczDo$ masses, the strongest exclusion is obtained if the slepton mass is close to the chargino mass.
In this case, excluded masses reach up to 1.4\TeV for $\PSQtDo$ and 900\GeV for $\PSGczDo$.
When the slepton mass is taken as the mean of the chargino and neutralino masses, these numbers decrease to 1.3\TeV for $\PSQtDo$ and 750\GeV for $\PSGczDo$.
A further reduction to 1.2\TeV for $\PSQtDo$ and to 100\GeV for $\PSGczDo$ is observed when the slepton mass is close to the neutralino mass.

\begin{acknowledgments}
  We congratulate our colleagues in the CERN accelerator departments for the excellent performance of the LHC and thank the technical and administrative staffs at CERN and at other CMS institutes for their contributions to the success of the CMS effort. In addition, we gratefully acknowledge the computing centers and personnel of the Worldwide LHC Computing Grid for delivering so effectively the computing infrastructure essential to our analyses. Finally, we acknowledge the enduring support for the construction and operation of the LHC and the CMS detector provided by the following funding agencies: BMBWF and FWF (Austria); FNRS and FWO (Belgium); CNPq, CAPES, FAPERJ, FAPERGS, and FAPESP (Brazil); MES (Bulgaria); CERN; CAS, MoST, and NSFC (China); COLCIENCIAS (Colombia); MSES and CSF (Croatia); RIF (Cyprus); SENESCYT (Ecuador); MoER, ERC IUT, PUT and ERDF (Estonia); Academy of Finland, MEC, and HIP (Finland); CEA and CNRS/IN2P3 (France); BMBF, DFG, and HGF (Germany); GSRT (Greece); NKFIA (Hungary); DAE and DST (India); IPM (Iran); SFI (Ireland); INFN (Italy); MSIP and NRF (Republic of Korea); MES (Latvia); LAS (Lithuania); MOE and UM (Malaysia); BUAP, CINVESTAV, CONACYT, LNS, SEP, and UASLP-FAI (Mexico); MOS (Montenegro); MBIE (New Zealand); PAEC (Pakistan); MSHE and NSC (Poland); FCT (Portugal); JINR (Dubna); MON, RosAtom, RAS, RFBR, and NRC KI (Russia); MESTD (Serbia); SEIDI, CPAN, PCTI, and FEDER (Spain); MOSTR (Sri Lanka); Swiss Funding Agencies (Switzerland); MST (Taipei); ThEPCenter, IPST, STAR, and NSTDA (Thailand); TUBITAK and TAEK (Turkey); NASU (Ukraine); STFC (United Kingdom); DOE and NSF (USA).

  \hyphenation{Rachada-pisek} Individuals have received support from the Marie-Curie program and the European Research Council and Horizon 2020 Grant, contract Nos.\ 675440, 752730, and 765710 (European Union); the Leventis Foundation; the A.P.\ Sloan Foundation; the Alexander von Humboldt Foundation; the Belgian Federal Science Policy Office; the Fonds pour la Formation \`a la Recherche dans l'Industrie et dans l'Agriculture (FRIA-Belgium); the Agentschap voor Innovatie door Wetenschap en Technologie (IWT-Belgium); the F.R.S.-FNRS and FWO (Belgium) under the ``Excellence of Science -- EOS" -- be.h project n.\ 30820817; the Beijing Municipal Science \& Technology Commission, No. Z191100007219010; the Ministry of Education, Youth and Sports (MEYS) of the Czech Republic; the Deutsche Forschungsgemeinschaft (DFG) under Germany's Excellence Strategy -- EXC 2121 ``Quantum Universe" -- 390833306; the Lend\"ulet (``Momentum") Program and the J\'anos Bolyai Research Scholarship of the Hungarian Academy of Sciences, the New National Excellence Program \'UNKP, the NKFIA research grants 123842, 123959, 124845, 124850, 125105, 128713, 128786, and 129058 (Hungary); the Council of Science and Industrial Research, India; the HOMING PLUS program of the Foundation for Polish Science, cofinanced from European Union, Regional Development Fund, the Mobility Plus program of the Ministry of Science and Higher Education, the National Science Center (Poland), contracts Harmonia 2014/14/M/ST2/00428, Opus 2014/13/B/ST2/02543, 2014/15/B/ST2/03998, and 2015/19/B/ST2/02861, Sonata-bis 2012/07/E/ST2/01406; the National Priorities Research Program by Qatar National Research Fund; the Ministry of Science and Higher Education, project no. 02.a03.21.0005 (Russia); the Programa Estatal de Fomento de la Investigaci{\'o}n Cient{\'i}fica y T{\'e}cnica de Excelencia Mar\'{\i}a de Maeztu, grant MDM-2015-0509 and the Programa Severo Ochoa del Principado de Asturias; the Thalis and Aristeia programs cofinanced by EU-ESF and the Greek NSRF; the Rachadapisek Sompot Fund for Postdoctoral Fellowship, Chulalongkorn University and the Chulalongkorn Academic into Its 2nd Century Project Advancement Project (Thailand); the Kavli Foundation; the Nvidia Corporation; the SuperMicro Corporation; the Welch Foundation, contract C-1845; and the Weston Havens Foundation (USA).
\end{acknowledgments}

\bibliography{auto_generated}
\cleardoublepage \appendix\section{The CMS Collaboration \label{app:collab}}\begin{sloppypar}\hyphenpenalty=5000\widowpenalty=500\clubpenalty=5000\vskip\cmsinstskip
\textbf{Yerevan Physics Institute, Yerevan, Armenia}\\*[0pt]
A.M.~Sirunyan$^{\textrm{\dag}}$, A.~Tumasyan
\vskip\cmsinstskip
\textbf{Institut f\"{u}r Hochenergiephysik, Wien, Austria}\\*[0pt]
W.~Adam, F.~Ambrogi, T.~Bergauer, M.~Dragicevic, J.~Er\"{o}, A.~Escalante~Del~Valle, R.~Fr\"{u}hwirth\cmsAuthorMark{1}, M.~Jeitler\cmsAuthorMark{1}, N.~Krammer, L.~Lechner, D.~Liko, T.~Madlener, I.~Mikulec, F.M.~Pitters, N.~Rad, J.~Schieck\cmsAuthorMark{1}, R.~Sch\"{o}fbeck, M.~Spanring, S.~Templ, W.~Waltenberger, C.-E.~Wulz\cmsAuthorMark{1}, M.~Zarucki
% ...
\vskip\cmsinstskip
\dag: Deceased\\
1:  Also at Vienna University of Technology, Vienna, Austria\\
% ...
\end{sloppypar}
\end{document}